\documentclass[12pt,tightenlines,eqsecnum,floats,showpacs,nofootinbib,amsmath,amssymb,aps,prd]{revtex4}

\usepackage{graphicx,verbatim}

\newcommand{\Lie}[0]{{\cal L}\,}
\newcommand{\grad}[0]{\nabla\!}
\newcommand{\pullback}[1]{\hbox{\lower0.5ex\hbox{${}_{\leftarrow}$}}\kern-1.9ex{#1}}

\def\man{\mathcal{M}}
\def\R{\mathcal{R}}

\def\t{\tilde}
\def\l{{\ell}}

\def\q{\tilde{q}}

\def\vp{\varphi}

\def\R{\mathcal{R}}

\def\lb{\bar{\ell}}
\def\nb{\bar{n}}

\def\th{{\widehat{\tau}}}
\def\rh{{\widehat{r}}}

\def\be{\begin{equation}}
\def\ee{\end{equation}}
\def\ba{\begin{eqnarray}}
\def\ea{\end{eqnarray}}

\def\tl{\Theta_{(\ell)}}
\def\tn{\Theta_{(n)}}
\def\tlb{\Theta_{(\bar{\ell})}}
\def\tnb{\Theta_{(\bar{n})}}
\def\w{\omega}
\def\sn{{\sigma}^{(n)}}
\def\snb{\sigma^{(\bar{n})}}
\def\sl{{\sigma}^{(\ell)}}

\def\eps{\epsilon}

\def\tt{\tilde{\tau}}

\def\f{\frac}

\def\qo{\mathring{q}}

\begin{document}

\title{Dynamical Black Holes: Approach to the Final State}
\author{Abhay Ashtekar}
\email{ashtekar@gravity.psu.edu} 
\author{Miguel Campiglia}
\email{miguel@rri.res.in} 
\author{Samir Shah}
\email{mail@sshah.org} \affiliation{Institute for Gravitation and the
Cosmos \& Physics Department, Penn State, University Park, PA 16802,
U.S.A.}

\begin{abstract}

Since black holes can be formed through widely varying processes,
the horizon structure is highly complicated in the dynamical phase.
Nonetheless,  as numerical simulations show, the final state appears
to be universal, well described by the Kerr geometry. How are all
these large and widely varying deviations from the Kerr horizon
washed out? To investigate this issue, we introduce a well-suited
notion of horizon multipole moments and equations governing their
dynamics, thereby providing a coordinate and slicing independent
framework to investigate the approach to equilibrium. In particular,
our flux formulas for multipoles can be used as analytical checks on
numerical simulations and, in turn, the simulations could be used to
fathom possible universalities in the way black holes approach their
final equilibrium.

\end{abstract}


\maketitle

\section{Introduction}

Black hole uniqueness theorems \cite{mh} strongly suggest that late
stages of a gravitational collapse or a black hole merger are well
described by the Kerr solution. In particular, once the black hole
reaches the final equilibrium, its horizon is expected to match the
Kerr isolated horizon which can be characterized intrinsically,
without reference to the exterior space-time \cite{lp}. By now a
wide variety of numerical simulations have confirmed this
expectation. However, these simulations also bring out the fact that
there is great diversity in the structure of the horizon during the
preceding dynamical phase. At its formation, the horizon of the
final black hole generically exhibits large, time-dependent
distortions. Heuristically, its intrinsic geometry appears to have
many `bumps' and there is no simple relation between its rotational
state and the spin vector of the final black hole. However, in the
process of settling down to equilibrium, the Einstein dynamics
manages to wash away these apparently large deviations leaving
behind the Kerr isolated horizon. How does this come about? Can one
provide a precise mathematical description of this approach to
equilibrium? Does it carry a clear imprint of general relativity
that could perhaps be seen in future gravitational wave
observations? The final state is universal. Are there universalities
associated also with the \emph{approach} to this final state?
Answers to these questions would provide us both a deeper conceptual
understanding of the \emph{strong field regime} of general
relativity and suggest avenues to test the theory through its
specific predictions for the \emph{non-linear, dynamical} phase of
black hole formation.

However, it is rather difficult to investigate these issues
precisely because the dynamical processes of interest occur in
the strong field regime of general relativity. Numerical
simulations have provided insights but the horizon distortions
seen in simulations often refer to components of geometrical
tensors in coordinate systems and, more importantly, foliations
they use. What one needs is an invariant characterization of
the horizon geometry in its dynamical phase. A natural avenue
is provided by the horizon multipole moments \cite{aepv,skb,ro}
which can be interpreted as the `source multipole moments' of
the black hole. However, as we discuss in sections \ref{s2.2}
and \ref{s4.3}, the current definitions are not as well-suited
to investigate the approach to equilibrium as one would like.

The purpose of this article is to provide multipole moments which
are well-tailored for this task and provide equations for their
dynamical evolution. These moments are just sets of numbers that
capture the diffeomorphism invariant content of dynamical and
arbitrarily distorted horizon geometries. Their evolution provides a
coordinate and slicing independent description of how black holes
shed the deviations from the Kerr horizon geometry and its spin
structure. These equations can be used as non-trivial checks on
numerical simulations in the strong field regime and, conversely,
numerical solutions of these equations will bring out universalities
in the approach to equilibrium, if they exist.

This article is organized as follows. In section \ref{s2} we collect
the material on isolated and dynamical horizons that serves as our
starting point. Main results are presented in sections \ref{s3} and
\ref{s4} which also include a discussion of the relation to other
definitions of multipoles in the literature \cite{skb,ro} and to
vortexes and tendexes that have been used to visualize the strong
field geometry near black holes \cite{vt1,vt2}. For the convenience
of computational relativists, in section \ref{s4} the ideas and
equations needed for numerical simulations have been presented in a
self-contained fashion. If the goal is only to use these multipoles
in numerical simulations, one can skip section \ref{s3} and go
directly to \ref{s4}. In section \ref{s5} we discuss their relation
to similar issues that have been explored in the literature,
including Price's law \cite{price1,price2,rd}, the close-limit
approximation \cite{pp,gnpp}, and the relation between dynamics at
the horizon and at infinity
\cite{eloisa,jaramillo1,jaramillo2,lr}.  
The Appendix collects a few analytical results on the the behavior
of the key fields on the dynamical horizon $H$ in the passage to
equilibrium: which of them diverge, which of them admit finite
limits and which of them vanish in the limit and at what rate.

Our conventions are as follows. We use Penrose's abstract index
notation. The space-time metric $g_{ab}$ has signature -,+,+,+ and
curvature tensors are defined by $2\nabla_{[a}\nabla_{b]} k_c =
R_{abc}{}^d k_d$, $R_{ac} = R_{abc}{}^b$ and $R = R_{ab}g^{ab}$.

\section{Quasi-local horizons}
\label{s2}

This section is divided into two parts. In the first we recall
the notions of isolated and dynamical horizons and their basic
properties\cite{ih-prl,abl1,ak}. In the second, we summarize
the definition of multipoles in the axi-symmetric case
\cite{aepv,skb}. These quasi-local horizons have had numerous
applications, including black hole thermodynamics
\cite{abl2,ak}, construction of initial data and extraction of
physics from numerical simulations
\cite{dkss,skb,akrev,gj,jlj}, and the definition of
\emph{quantum} horizons and analysis of their properties
\cite{abk,aev} in loop quantum gravity.

\subsection{Dynamical and isolated horizons}
\label{s2.1}

The notion of event horizons has played a major role in the
discussion of black holes. However, it is teleological and `too
global' in that one needs the entire space-time evolution before one
can locate it. Dynamical and isolated horizons are quasi-local
notions which are free from these limitations.%
\footnote{Since our goal is only to convey the main ideas, the
discussion will be brief and we will have to gloss over some
finer points. For details and precise statements of results and
properties, see \cite{abl1,ak,ag,akrev,gj,jlj}.}

\begin{quote}
A \emph{dynamical horizon} (DH) $H$ is a 3-dimensional space-like
sub-manifold (possibly with boundary) of space-time $(\man,
g_{ab})$, foliated by a family of 2-spheres $S$ such that:\\ i) Each
$S$ is marginally trapped; i.e. the expansion $\tl$ of one of the
(future directed) null normals $\ell^a$ to each $S$ vanishes, and,\\
ii) The expansion $\tn$ of the other (future directed) null normal
is negative.
\end{quote}
Heuristically, since $H$ is obtained by `stacking together'
marginally trapped surfaces (MTSs), it can be thought of as the
boundary of a trapped region of space-time representing a black
hole. The area of the MTSs $S$ increases in time, depicting a
\emph{dynamical phase} during which the black hole grows as it
swallows matter and gravitational waves. Furthermore, Einstein's
equations imply that there is a detailed balance law equating the
rate of growth of the area-radius $R_{S}$ of any MTS $S$ with the
total flux of energy (in matter and gravitational waves) falling
into the black hole across $S$ \cite{ak}.

Given a DH $H$, one can show that it does not admit any MTS that is
not in the foliation. Thus, the foliation by MTSs ---the `internal
structure' of $H$--- is unique. DHs naturally arise in numerical
simulations where one begins with a foliation of space-time and uses
efficient algorithms to zero-in on the outermost MTSs. A local
existence theorem ensures that, given such an MTS, it will `evolve'
to a DH (provided certain generic conditions are met)
\cite{ams,amms}. However, DHs are not unique: a space-time region
that appears to represent a black hole can carry multiple DHs.
Nonetheless, partial uniqueness theorems do exist. In particular
they imply that in the numerical relativity constructions, there is
a unique DH that asymptotes to the event horizon in the distant
future \cite{ag}. This is the situation of interest in this paper.

Once the flux of energy across the horizon becomes zero, the horizon
becomes isolated. More precisely:
\begin{quote}
An \emph{isolated horizon} (IH) $\Delta$ is a null, 3-dimensional
sub-manifold in $(\man, g_{ab})$, topologically $S^2\times R$ and
equipped with a specific null normal $\lb^a$ such that:\\
i) The expansion $\tlb$ of $\lb^a$ vanishes;\\
ii) $\mathcal{L}_{\lb} q_{ab} =0$; and, \\
iii) $(\mathcal{L}_{\lb} D_a - D_a \mathcal{L}_{\lb})t^a =0$.
\end{quote}
Here $q_{ab}$ is the intrinsic (degenerate) metric on $\Delta$, $D$
the derivative operator induced on $\Delta$ by the space-time
derivative operator $\nabla$, and $t^b$ is any vector field that is
tangential to $\Delta$.%
\footnote{Note that $q_{ab}$ has signature 0,+,+ with $\lb^a$ as the
degenerate direction; $q_{ab}\lb^b =0$. Condition ii) implies that
$\nabla$ induces a well-defined derivative operator $D$ on $\Delta$.
It is automatically satisfied if the stress-energy tensor satisfies
a mild version of the dominant energy condition:
$-T^{a}{}_{b}\ell^b$ is a future directed causal vector everywhere
on $\Delta$.}

The fields $(q_{ab}, D)$ constitute the \emph{intrinsic
geometry} of the IH $\Delta$. By requiring that $(q_{ab}, D)$
be time-independent (with respect to the evolution defined by
$\lb^a$), the notion of an IH extracts from that of Killing
horizons just the minimal properties to ensure that the horizon
itself is in equilibrium, allowing for dynamical processes to
occur arbitrarily close to it \cite{pc,akrev}. The definition
ensures that neither matter nor gravitational waves fall across
$\Delta$ and the area of any 2-sphere cross section of $\Delta$
is the same. Event horizons of stationary black holes are
simplest examples of IHs \cite{ih-prl,abl1,akrev,gj,jlj}.

Consider formation of a black hole via gravitational collapse or
merger of two compact objects, one or both of which may be black
holes. We are primarily interested in the late stage of such
processes, when a common DH $H$ develops and approaches an IH
$\Delta$ representing the future part of the event horizon of the
final black hole. Because of back-scattering of gravitational waves,
in the exact theory the approach would only be asymptotic. However,
in numerical simulations one invariably finds that the back
scattering becomes negligible within numerical errors rather soon
and $H$ joins on to $\Delta$ at some finite time. Therefore, in this
paper we will focus on this situation. (The case in which the
equilibrium is reached only asymptotically is in fact somewhat
simpler \cite{ak,bf}.)

\subsection{Mutipole moments: The axi-symmetric case}
\label{s2.2}

Numerical simulations invariably use convenient choices of
coordinates and foliations and these choices vary from one research
group to another. Therefore, the task of comparing the final results
requires analytical tools to probe and compare distinct horizon
geometries in an invariant fashion. Multipole moments provide such a
tool. In this sub-section we will summarize the situation in the
case when the horizons are axi-symmetric \cite{aepv,skb,akrev}.

Let us begin with IHs $\Delta$. An IH $\Delta$ is said to be
\emph{axi-symmetric} if it admits a vector field $\vp^a$ satisfying:
$\mathcal{L}_\vp \lb^a = 0,\,\, \mathcal{L}_\vp q_{ab} =0$, and
$(\mathcal{L}_{\vp} D_a - D_a \mathcal{L}_{\vp}) t^b = 0$ for all
vectors $t^a$ tangential to $\Delta$. Thus, diffeomorphisms
generated by $\vp^a$ on $\Delta$ preserve its geometry. These
conditions imply that $\vp^a$ has an unambiguous projection on the
2-sphere of integral curves of $\lb^a$ which is a rotational Killing
field there.

Now, it is known that the diffeomorphism invariant content of the
geometry $(q_{ab}, D)$ of $\Delta$ is captured in two fields:\\
i) The scalar curvature $\R$ of $\q_{ab}$, the induced metric on any
2-sphere cross-section $S$ of $\Delta$, and,\\
ii) the `rotational' 1-form $\omega_a$ on $\Delta$ defined by $D_a
\lb^a =\omega_a \ell^a$ \cite{abl1,akrev}.\\
The geometrical relation of these fields is brought out by the Weyl
tensor. On any IH, the component $\Psi_2$ of Weyl curvature is gauge
invariant and furthermore we have:
\be \label{weyl} \Psi_2 = \textstyle{\f{1}{4}}\, \R\, +\,
\textstyle{\f{i}{2}} {\epsilon}^{ab} D_a \omega_b .\ee
Here ${\epsilon}^{ab}$ is the area bi-vector on any 2-sphere cross
section of $\Delta$ (and the right hand side is independent of the
specific choice of the cross-section $S$). Thus, on $\Delta$, the
scalar curvature $\R$ is essentially the same as the real part of
$\Psi_2$ while the rotational 1-form is a potential for the
imaginary part of $\Psi_2$. In numerical simulations, one can
calculate these fields on $\Delta$. However, it is still not
possible to compare the results of two different simulations because
the fields live on two different 3-manifolds $\Delta$ and there is
no natural identification between them. Geometric multipoles are two
sets of numbers $I_l, L_l$, with $l = 0,1,2, \ldots$ which capture
the entire diffeomorphism invariant content of these fields
\cite{aepv}. Therefore, to compare the results of any two
simulations, it suffices to compute these numbers in each simulation
and compare them. In practice, it suffices to compare just the first
few multipoles.

The key idea behind the definition of multipoles is the following.
Given an axi-symmetric metric $q_{ab}$ on a 2-sphere $S$, one can
construct a \emph{canonical} round 2-sphere metric $\qo_{ab}$ on $S$
together with a preferred rotational Killing field \cite{aepv}. This
structure in turn provides canonical weighting functions $Y_{l, m}$,
the spherical harmonics of $\qo_{ab}$. The multipoles are now
defined as:
\ba \label{mm1} I_{l,m} - iL_{l,m} &:=&  \oint_S
[\textstyle{\f{1}{4}}\, \R\, +\, \textstyle{\f{i}{2}} \epsilon^{ab}
D_a \omega_b]\,Y_{l, m}\, d^2V\\  &\equiv& \oint_S \Psi_2\, Y_{l,
m}\, d^2V\, \label{mmih} .\ea
where the integral is performed on \emph{any} 2-sphere cross-section
$S$ of $\Delta$ and $d^2V$ is the volume element on $S$. Of course,
because the horizon geometry is axi-symmetric, only the $m=0$
multipole moments are non-vanishing. Furthermore, $I_{0,0}$ is just
$1/4$th the Gauss invariant, $I_{0,0} = 2\pi$, and $L_{0,0}$
vanishes. Therefore only the $l \not= 0$ moments are non-trivial.

Since each step in the construction is diffeomorphism covariant
---none involved introduction of a structure other than the
given axi-symmetric IH--- the final numbers are diffeomorphism
invariant. A given axi-symmetric horizon geometry yields these
numbers and, conversely, given the numbers that arise from an
axi-symmetric horizon geometry, one can reconstruct that geometry up
to an overall diffeomorphism. Finally, by a simple rescaling of
these geometrical multipoles, one can obtain the \emph{mass and spin
multipoles} associated with the horizon. Since these refer only to
the horizon without any reference to the exterior space-time region,
they represent the \emph{source multipoles} associated with the
black hole itself. Indeed, as explained in \cite{aepv}, the
construction suggests that one can assign a `surface mass density'
$\rho_\Delta = -(1/8\pi) M_\Delta\, \R$ and a `surface spin current'
$j^{\Delta}_a = (1/8\pi G) \omega_a$ to the isolated horizon
$\Delta$, where $M_\Delta$ is the total mass of $M_\Delta$. By
contrast, the multipole moments defined at infinity represent `field
multipoles' which include contributions not only from the black hole
but also from the exterior gravitational field (and matter, if any).
In the Newtonian theory, the two sets agree. But because of its
non-Abelian character, in general relativity gravity sources
gravity. Therefore the two moments differ. For the mass quadrupole
in Kerr space-time, for example, the difference increases with spin
and is of the order of 40\% near extremality $a\sim m$ \cite{aepv}.

What about dynamical horizons $H$? The diffeomorphism invariant
content of the intrinsic geometry of any MTS $S$ is again encoded in
the scalar curvature $\R$ of $S$, while the role played by the
rotational 1-form is now played by $\t{\omega}_a := {\q}_a{}^b
K_{bc} \rh^{\,c}$ where $\q_{ab}$ is the intrinsic 2-metric on $S$
and $\rh^{\,c}$ is the unit (space-like) normal to $S$ within $H$
and $K_{ab}$ is the extrinsic curvature of $H$ in space-time.%
\footnote{This follows from the following considerations involving
the `Weingarten map'. On an IH $\Delta$, the 1-form $\omega_a$ that
features in (\ref{mm1}) is the pull-back to a 2-sphere cross-section
$S$ of $\Delta$ of the one-form $- (1/2)\,\nb_b\, D_a \lb^b \equiv -
(1/2) \nb_b\,\, \pullback{\nabla_a}{\lb}^b$ where
$\pullback{\nabla_a}$ is the pull-back to $\Delta$ of the space-time
connection. On a DH, the 1-form $\t{\omega}_a := \q_{a}{}^b
K_{bc}\rh^{\,c}$ is given by the pull-back to MTSs $S$ of $-
(1/2)\,n_b\, \pullback{\nabla_a}{\th}^{\,b}$ where
$\pullback{\nabla_a}$ is the pull-back to $H$ of the space-time
connection and $\th^{\,b}$ is the unit time-like normal to $S$. As
in \cite{ak}, we use the conventions $\ell^a n_a = -2 =
\lb^a\nb_a$.}
Using the same motivation as on IHs, one can introduce an effective
`mass surface density' and an `angular spin current density' on any
MTS $S$ of the DH $H$ and they are given by $\rho_S = -(1/8\pi)
M_S\, \R$ and $j^{S}_a = (1/8\pi G) \t{\omega}_a$, where $\R$ is the
scalar curvature of the 2-metric $\t{q}_{ab}$ on $S$
\cite{akrev,skb}. Therefore, in the numerical relativity literature
the definition (\ref{mm1}) has been recast in terms of these fields,
\be \label{mm2} I_{l,m}[S] - iL_{l,m}[S] :=  \oint_S
[\textstyle{\f{1}{4}}\, \R\, +\, \textstyle{\f{i}{2}} \epsilon^{ab}
\t{D}_a \t{\omega}_b]\,\,Y_{l, m}\, d^2V, \quad {\rm where} \quad
\t{\omega}_a = \q_{a}{}^b K_{bc} \rh^{\,c}\, ,\ee
and taken over to assign multipole moments $I_{l,0},\, L_{l,0}$ with
each marginally trapped surface $S$ in the foliation \cite{skb}.
(Note that, whenever there is possible ambiguity, we use tilde over
symbols that refer to 2-dimensional fields on the MTSs.)

On a DH, these multipole moments change in time, capturing the
`intrinsic' dynamics of the black hole, encapsulated in the horizon
geometry. However, to implement this strategy, one has to find an
axial symmetry $\vp^a$ on each $S$. There \emph{are} efficient
numerical algorithms to locate this required axial Killing field
$\vp^a$, if it exists \cite{dkss,kvf1,kvf2,kvf3,kvf4}. However, as
one might expect, the DH formed in a gravitational collapse or a
black hole merger generically fails to be even approximately
axi-symmetric except at very late time when the geometry is already
close to that of the Kerr IH. Therefore the strategy is not
well-suited to study how the horizon loses its `hair' in its
approach to the final Kerr state. Indeed, in the dynamical phase one
expects the black hole spin, for example, to change not only in
magnitude but also in direction, while the axi-symmetry assumption
forces the angular $l=1$ momentum moment to have only the
`z-component'. More generally, one would expect most moments to have
non-zero values for $m\not=0$ and it is of significant interest to
see how dynamics of general relativity forces the black hole to shed
them as it approaches equilibrium. To probe this issue, in sections
\ref{s3} and \ref{s4} we will generalize the framework by going
beyond axi-symmetry in a manner that is well-suited to understanding
the passage to equilibrium. We will also comment on the relation of
this strategy to another approach \cite{ro}
that has been proposed in the literature.\\

\emph{Remarks:}

1. In recent years, there has been considerable interest in using
the Kerr multipoles to test the no-hair theorems of general
relativity through gravitational wave signals. Much of this analysis
is based on some key ideas introduced by Ryan \cite{fr} using
signals arising from a compact object orbiting around a supermassive
black hole. The strategy is to express the metric of the
supermassive black hole at the location of the compact object as an
expansion, with the Geroch-Hansen field multipoles at infinity as
coefficients \cite{rg,rh,bs}. However, it would seem that the
expansion of the space-time metric in terms of the source multipoles
that characterize the horizon geometry would provide a more accurate
route to mapping the Kerr geometry, unless the orbiting compact
object is truly in the asymptotic region, very far from the central
black hole. If it is closer, then expanding the space-time metric
`outward' starting from the horizon \cite{ih-prl}, rather than
`inward' from infinity, should require far fewer terms to attain the
desired accuracy. There is also a conceptual advantage that one
would only need to assume vacuum equations in the region between the
two bodies.

2. The simple relation (\ref{weyl}) between the fields $\R$ and
$\omega_a$ and the Weyl curvature component $\Psi_2$ on IHs is
modified on a DH. We now have
\ba  \R &=& \left(4\, {\rm Re} \Psi_2  -\t{q}^{ab} \t{q}^{cd}
{\sl}_{ac} \sn_{bd}\right)\\
2\, {\eps}^{ab} \t{D}_a \t{\omega}_b &=&\left(4\,{\rm Im} \Psi_2 +
{\eps}^{ab}\t{q}^{cd}\sl_{ac}\sn_{bd}\right) \, , \ea
where $\sn_{ab}$ and $\sl_{ab}$ are the shears associated with the
null normals $\ell^a$ and $n^a$ to the MTSs $S$ and $\t{q}^{ab}$ is
the metric on $S$. Therefore, on a DH, multipoles are no longer
determined by $\Psi_2$ alone. (When the horizon becomes isolated,
$\sl_{ab}$ vanishes and the extra term drops out.)


\section{Multipole moments of general quasi-local horizons}
\label{s3}

In this section we present the conceptual strategy which allows us
to define multipole moments on general, non-axi-symmetric horizons
and track their time evolution. The material is divided into four
parts. In the first, we introduce the main idea behind the
generalization to non-axi-symmetric contexts; in the second, we
execute this strategy, in the third, we present the generalized
multipoles and, in the fourth, we present `balance laws' that
dictate the dynamics of multipole moments.

\subsection{Main ideas}
\label{s3.1}

The underlying strategy is the same for both sets of geometric
moments $I_{l,m}$ and $L_{l,m}$. We will first describe it in detail
for the geometric spin moments $L_{l,m}$ and then summarize the
situation for the $I_{l,m}$. In the first part of the discussion, we
will consider the isolated and dynamical horizons simultaneously.
For IHs, $S$ can be any cross section (or the 2-sphere of the null
generators $\bar{\ell}^a$) of $\Delta$ while for DHs, $S$ can be any
MTS.

Let us first integrate the expression (\ref{mm2}) for $L_{l,m}$ by
parts to obtain
\be \label{amm} L_{l,m}[S] = - {\textstyle{\f{1}{2}}
\sqrt{\f{2l+1}{4\pi}}\, R^{-2}}\,\, \oint_S \vp^a_{l,m}\,
\t{\omega}_a\, d^2V\quad {\rm where} \quad \vp_{l,m}^a =
\textstyle{\sqrt{\f{4\pi}{2l+1}}\, R^2\,} {\epsilon}^{ab} D_b
Y_{l,m}\,. \ee
where, as before, $R$ is the area-radius of $S$ and we have
introduced certain normalization factors for later convenience. Note
that the $\vp^a_{l,m}$ are all divergence-free on $S$ and,
furthermore, they  provide a complete basis on the space of
divergence-free vectors. Therefore $L_{l,m}$ can be thought of as
providing a linear map from a basis of divergence-free vector fields
on $S$ to reals. In this respect, there is a structural similarity
between multipole moments on $\Delta$ or $H$ and `conserved' charges
at null infinity, which can be regarded as linear maps from the
generators of the Bondi-Metzner-Sachs (BMS) group to the reals
\cite{scri0,scri1,scri2,scri3}. With multipoles, the divergence-free
vector fields play the role of infinitesimal symmetries. This
conceptual parallel will be useful in our discussion.

In the axi-symmetric case, we have a symmetry vector field $\vp^a$
and only $L_{l,0}$ are non-zero. In the language of vector fields
these correspond to moments associated with the $\vp_{l,0}^a$
satisfying ${\cal L}_\vp\, \vp^a_{l,0} = 0$. In the literature one
often sets $Y_{1,0} = \sqrt{3/4\pi}\,\zeta$. Then $Y_{l,0}$ are all
essentially just the Legendre polynomials in $\zeta$;\, $Y_{l,0} =
\sqrt{(2l+1)/4\pi}\,\, P_l (\zeta)$. The function $\zeta$ is singled
out by the axial Killing field: $\vp^a = R^2\, \epsilon^{ab} \t{D}_b
\zeta\, \equiv \, \vp^a_{1,0}$ (whence $\vp^a = \vp^a_{1,0}$). On a
general horizon, the major obstacle has been that we do not have
access to this route; without axi-symmetry, there is no preferred
$\zeta$ on $S$ and hence we do not have the required basis $Y_{l,m}$
of functions.

The first step in the generalization is just to forego the preferred
basis and use (\ref{amm}) to associate multipole moments $L_\phi$
with \emph{any} divergence-free vector field $\phi^a$ on $S$:
\be \label{L} L_{\phi} [S] = - {\textstyle{\f{1}{2}}}\, \oint_S
\phi^a\, \t{\omega}_a\, d^2V\quad {\rm where} \quad {\cal L}_\phi\,
\epsilon^{ab} = 0. \ee
But since the vector fields $\phi^a$ are defined separately on each
$S$, we need a prescription to identify vector fields that lie on
\emph{different} cross-sections $S$. Otherwise, we would not be able
to compare multipoles associated with two different cross-sections:
On an IH the definition would be ambiguous and on a DH we would not
be able to study the evolution of multipoles.

On IHs the required identification is easy to achieve: consider the
diffeomorphism generated by the appropriate (possibly angle
dependent) multiple of the vector field $\lb^a$ that maps the first
cross-section $S_1$ to the second $S_2$. This natural map --the
analog of the BMS super-translation at null infinity-- sends
divergence-free vector fields on $S_1$ to divergence-free vector
fields on $S_2$. With this identification between divergence-free
vector fields, it follows that multipole moments are independent of
the choice of the cross-section $S$. Equivalently, we can use the
2-sphere of generators $\lb^a$ of $\Delta$ for $S$ in (\ref{L}).
This simpler procedure makes it manifest that the multipoles
$L_\phi$ are properties of the IH as a whole.

On DHs, on the other hand, the geometry and hence the multipoles
evolve in time and we need to follow the analog of the first
procedure. Now $S$ can be any one of the MTSs. Therefore, we need to
construct a \emph{dynamical vector field} $X^a$ on $H$ that provides
a natural identification between the leaves of the foliation
provided by MTSs. Motions along $X^a$ will then be interpreted as
`time evolution'. We need this vector field $X^a$ to have the
following four properties:

\begin{itemize}
\item i) The 1-parameter family of diffeomorphisms generated by
    $X^a$ on $H$ should preserve the foliation by MTSs;
\item ii) It should provide an isomorphism between the space of
    divergence-free vector fields on any $S$ to that of
    divergence-free vector fields on its image;
\item iii) $X^a$ should be constructed covariantly, using only
    that structure which is already available on general
    dynamical horizons without any symmetry; and,
\item iv) If the DH is axi-symmetric, diffeomorphisms generated
    by $X^a$ should preserve the symmetry vector field
    $\varphi^a$. As we will see this will guarantee that the
    multipole moments given by the more general construction
    --that does not refer to axi-symmetry at all-- do reduce
    to the multipoles used in the literature in the
    axi-symmetric case \cite{skb}.
\end{itemize}
We will show that one can select, in a diffeomorphism covariant
fashion, a class of vectors fields $X^a$ satisfying these properties
on any DH and multipoles are insensitive to the choice of $X^a$
within this class.

\subsection{Determining the dynamical vector field $X^a$}
\label{s3.2}

Since we already have a natural foliation by MTSs, any dynamical
vector field $X^a$ on $H$ can be decomposed into a part that is
orthogonal to the foliation and a part that is tangential: $X^a = N
\rh^{\,a} + N^a$ where, as before, $\rh^{\,a}$ is the unit normal to
each leaf of the foliation. Because $X^a$ must map every MTS to some
other MTS, the `lapse' $N$ is severely restricted. To write out the
restriction explicitly, let us introduce a coordinate $v$ on $H$
such that the leaves of the foliation are given by $v = {\rm
const}$. Then $N = C (q^{ab} D_a v D_b v)^{-1/2}$ where $C$ is a
constant and $q^{ab}$ the inverse of the intrinsic +,+,+ metric
$q_{ab}$ on $H$. Without loss of generality, we can set $C=1$ making
$v$ the affine parameter of the vector field $X^a$. This choice of
`lapse' is denoted by $2b$ in the literature. Thus, we have
\be\label {X}  X^a = 2b\, \rh^{\,a} + N^a\quad {\rm where} \quad 2b
= |Dv|^{-\textstyle\f{1}{2}} \, ,\ee
and it now remains to determine the `shift' $N^a$. We will now show
that the shift is also naturally fixed by our requirements. 

We will first describe the strategy. The dynamical horizon is
naturally equipped with a 2-form $\epsilon_{ab}$ that serves as the
area 2-form on each MTS: $\epsilon_{ab}= \epsilon_{abc}\,\rh^{\,c}$
where $\epsilon_{abc}$ is the volume 3-form on $H$. Had
$\mathcal{L}_{2b\hat{r}}\,\epsilon_{ab}$ been zero, $2b\rh^{\,a}$
would have mapped divergence-free vector fields on any $S$ to
divergence-free vector fields on its image and we could just set
$X^a = 2b\,\rh^{\,a}$, i.e. choose the shift $N^a$ to be zero. But
this strategy is not viable because
$\mathcal{L}_{2b\hat{r}}\,\epsilon_{ab} = 2b\,\t{K}\,
\epsilon_{ab}$, where $\t{K}$ is the trace of the extrinsic
curvature of the MTS $S$ within $H$ which is necessarily non-zero
because the area of these MTSs increases in time. The idea is to
remedy this `problem' with an appropriate choice of the shift $N^a$.
But no matter which shift $N^a$ we use, we will not be able to
compensate for the entire term $2b\,\t{K}$: Since $\mathcal{L}_N \,
\epsilon_{ab} = (\t{D}_aN^a)\, \epsilon_{ab}$, even with a judicious
choice of the shift $N^a$, we can only remove the \emph{purely
inhomogeneous} part
\be 2b\, \t{K} - (1/4\pi R^2)\, \oint_S 2b\, \t{K} d^2V\, =\,\, 2b\,
\t{K} - 2\dot{R}/R \ee
of $2b\, \t{K}$, where the area of each MTS is given by $4\pi R^2$,
and the `dot' denotes the derivative w.r.t. $v$. Then, although
$\mathcal{L}_X \epsilon_{ab}$ will not vanish, it will be of the
form $f(v) \epsilon_{ab}$, where $f(v) = 2\dot{R}/R$. Clearly, this
is the best one can hope for, given the fact that the area of the
MTSs changes with  $v$. But since $v$ is constant on any $S$, this
is sufficient to guarantee that the diffeomorphisms generated by
$X^a$ will map divergence-free vector fields on any MTS $S$ to
divergence-free vector fields on its image.

Let us now implement this strategy. First, we construct a unique
function $g$ on each $S$ such that:
\be \label{g}\t{q}^{ab} \t{D}_a \t{D}_b g = -(2b \t{K} -
2{\dot{R}}/{R}) \quad\quad {\rm and} \quad\quad \oint_S \, g\, d^2 V
= 0\ee
where, as before, the tilde quantities refer to the intrinsic
2-geometry of each MTS. The existence of the solution to the first
equation is guaranteed because its right hand side integrates out to
zero and the second equation makes the solution unique by removing
the freedom to add a constant to $g$. We then set
\be \label{shift1} N^a = \t{q}^{ab} \t{D}_b g, \quad\quad \hbox{\rm so
that}\quad \t{D}_aN^a = -(2b\t{K} - 2{\dot{R}}/{R})\,  \ee
so that $X^a = 2b\, \rh^{\,a} + N^a$ satisfies $\mathcal{L}_X \,
\epsilon_{ab} = (2\dot{R}/R)\, \epsilon _{ab}$, or, $\mathcal{L}_X
\, R^{-2}\, \epsilon_{ab}=0$. Note that, since $\tilde{K}= -(1/2)b\,
 \tnb$, and $\bar{n}^a$ is smooth on all of $M$,
$b\,\tilde{K}$ vanishes in the limit $v\to v_o$. Since $\dot{R}$
also vanishes, it follows from (\ref{g}) that $g$ and hence the
shift $N^a$ vanishes on $S_o$ and $X^a$ joins on smoothly with
$\bar\ell^a$ there.

By its construction, $X^a$  satisfies the first three of our four
requirements: It is constructed covariantly, preserves the
foliation, maps divergence-free vector fields on any $S$ to
divergence-free vector fields on its image. It turns out that it
also satisfies the fourth requirement. To see this, let us suppose
the DH is axi-symmetric with an axial symmetry vector field $\vp^a$.
Then, since our construction of $X^a$ uses only the horizon
geometry, it follows that $\mathcal{L}_\vp X^a = 0$. Therefore the
diffeomorphisms generated by $X^a$ map the axi-symmetry vector field
$\vp^a$ on any given MTS $S$ to the axi-symmetry vector field
$\vp^a$ on its image. We will see in section \ref{s3.3} that this
implies that, in the axi-symmetric case,  the multipole defined
using this general strategy coincide with those defined using
axi-symmetry as in \cite{skb}.

Finally, what would happen if we replace the coordinate $v$ labeling
the MTSs by $v' = f(v)$, where $f$ is a monotonic function of $v$?
It is straightforward to check that $X^a \mapsto X^\prime{}^a =
\dot{f}^{-1}X^a$. A vector field $\phi^a$ which is everywhere
tangential to the MTSs and divergence-free on them satisfies
$\mathcal{L}_X \phi^a =0$ if and only if it satisfies
$\mathcal{L}_{X^\prime} \phi^a =0$. Therefore, the `permissible'
divergence-free vector fields $\phi^a$ selected by $X^a$ are the
same as those selected by $X^\prime{}^a$, whence the multipoles
$L_\phi [S]$ of Eq (\ref{L}) are also the same.

\subsection{Generalized multipoles}
\label{s3.3}

We can now readily combine the results of the last two subsections
to define the generalized geometric spin multipoles. We first
introduce a vector field $X^a = 2b\rh^{\,a} + N^a$ on $H$ where $2b$
is given by (\ref{X}) and $N^a$ by (\ref{shift1}) and (\ref{g}).
Using it, we can single out the \emph{admissible} weighting fields
$\phi^a$: A vector field $\phi^a$ on $H$ which is tangential to
every MTS, and divergence-free on it, is an admissible weighting
field if $\mathcal{L}_X \phi^a =0$. Note that every admissible
vector field can be obtained simply by fixing a MTS $\bar{S}$, and a
divergence-free vector field $\bar{\phi}^a$ thereon, and Lie
dragging it along $X^a$. Given an admissible weighting field
$\phi^a$ and a MTS $S$, we now define the spin multipole moments
$L_\phi[S]$ following Eq. (\ref{L}):
\be \label{L2} L_{\phi} [{S}] = - {\textstyle{\f{1}{2}}}\,
\oint_{{S}} \phi^a\, \t{\omega}_a\, d^2V\, . \ee
By varying $S$ we can study the dynamical evolution of these
multipoles. Our weighting fields $\phi^a$ are `time independent' in
the sense that $\mathcal{L}_X \, \phi^a =0$. Therefore the multipole
moments $L_{\phi} [S]$ derive their time dependence solely from the
time dependence of the horizon geometry encoded in $\t\omega_a$ and
the 2-sphere volume element.

Next, let us discuss the extension of the second set of multipoles,
$I_{l,m}$, from axi-symmetric horizons to generic ones. For this we
first note that any metric 2-sphere $S$ admits an Abelian $U(1)$
connection $\Gamma_a$ whose curvature 2-form is determined by the
scalar curvature $\R$ of the metric:  $\t{D}_{[a} \Gamma_{b]} =
(\R/4) \epsilon_{ab}$ where $\epsilon_{ab}$ the area 2-form of $S$.
Therefore, one can think of repeating the above procedure, and
defining the other set of multipoles simply by replacing the 1-form
$\t{\omega}_a$ by $\Gamma_a$ in (\ref{L}).

However, there is a subtlety. While $\t{\omega}_a$ is defined
globally on the 2-spheres $S$, the connection 1-form $\Gamma_a$ has
to be defined in patches: Since $\oint_S \t{D}_{[a} \Gamma_{b]}\,
\epsilon^{ab}\, d^2V = (1/2)\oint_S \R\, d^2V = 4\pi$, the connection
1-form is globally defined only on the non-trivial $U(1)$ bundle
over $S^2$ with the
first Chern class.%
\footnote{In the language that is more familiar in the numerical
relativity literature, we have a connection that acts on the complex
dyad $m^a, \bar{m}^a$ on $S$ which is orthonormal in the sense $
\t{q}_{ab}m^a m^b =0$ and $\t{q}_{ab} m^a {\bar{m}}^b = 1$: $\t{D}_a
m_b = i\Gamma_a m_b$. The $U(1)$ gauge freedom corresponds to the
local rotations of the dyad via $m^a \mapsto e^{i\theta} m^a$ where
$\theta$ is a function on $S$. Neither the dyad nor the connection
is globally defined on $S$. But we can define them in patches and in
the overlap region the two sets are related by a gauge
transformation, $m^\prime{}^a = e^{i\theta} m^a$ and
$\Gamma^\prime{}_a = \Gamma_a - i\t{D}_a \theta$.}
But we can just fix a fiducial connection $\mathring{\Gamma}_a$
which is compatible with a round 2-sphere metric $\qo_{ab}$ whose
area 2-form is the same as that of the given physical metric
$q_{ab}$ on $S$. Then $C_a := \Gamma_a - \mathring{\Gamma}_a$
\emph{is globally defined} on $S$ with the property that $\t{D}_{[a}
C_{b]} = (1/4) (\R-\mathring{\R}) \epsilon_{ab}$, where
$\mathring{\R} = 2/R^2$ where $R$ is the area radius of $S$. Then,
for each permissible vector field $\phi^a$ on any MTS $S$ of the
horizon, we can set
\be \label{I} I_\phi [S] = {\textstyle{\f{1}{2}}}\, \oint_S \phi^a
C_a \, d^2 V   \quad\quad \hbox{\rm for any $\phi^a$ such that}\quad
\mathcal{L}_\phi\, \epsilon_{ab} =0. \ee
Although $\mathring\Gamma_a$ is arbitrary, because each $\phi^a$ is
divergence-free, the integral is in fact independent of the choice
of $\mathring\Gamma_a$ because $\oint_S \R d^2V = \oint_S
\mathring{\R} d^2V =8\pi$, the Gauss invariant of a 2-sphere.

Let us summarize. Given a generic DH $H$ we have introduced a family of
vector fields $X^a$, unique up to a rescaling by a function that is
constant on each MTS. The definition of this family is covariant and
constructive: Given \emph{any} DH, one can construct this family
using only the structure that is already available. The
diffeomorphisms generated by any of these $X^a$ preserve the
foliation by MTSs. We then defined \emph{permissible weighting
fields} $\phi^a$ on $H$; each $\phi^a$ is `time-independent',
tangential to each MTS and divergence-free on it. \emph{This family
of $\phi^a$ refers only to the geometric structure that is naturally
available on $H$.} They generalize the weighting functions $Y_{l,m}$
used on axi-symmetric horizons. Given a permissible weighting field
$\phi^a$, we use (\ref{L}) and (\ref{I}) to define geometric
multipoles $I_{\phi}[S],\, L_{\phi} [S]$ on any MTS $S$. By varying
$S$ we track its time development.

What if the DH under consideration is axi-symmetric? Then, as we saw
in section \ref{s3.2}, the axial symmetry field $\vp^a$ is
guaranteed to be `time independent', i.e., Lie dragged by $X^a$.
Now, by construction, $\mathcal{L}_X\, R^2 \epsilon^{ab} =0$ and,
since $\vp^a = R^2\, \epsilon^{ab} \t{D}_b \zeta$ with $\zeta$
satisfying $\oint_S \zeta\, d^2V =0$, it follows that
$\mathcal{L}_X\, \zeta =0$. Therefore, the vector fields
$\vp^a_{l,0} := R^2 \epsilon^{ab}\, \t{D}_b\, P_{l}(\zeta)$ are all
permissible in our general setting. In this setting, they define
multipoles via (\ref{I}) and (\ref{L2}). From (\ref{L}) it is clear
that this general definition agrees with the definition introduced
in \cite{skb}. Put differently, in the axi-symmetric case, the
function $\zeta$ defined separately on each cross-section using the
axial symmetry field $\vp^a$ is automatically `time independent',
i.e. satisfies $\mathcal{L}_{X}\, \zeta =0$ in the language of our
general setting. Therefore, with the identification $\phi^a =
\vp^a_{l,0}$, the multipoles $L_{l,0} [S]$ defined in the
axi-symmetric case (\ref{mm2}) coincide with the multipoles
$L_\phi[S]$ defined by the more general procedure, that does not
refer to axi-symmetry at all.

\subsection{Balance laws}
\label{s3.4}

On the DH, we have balance laws which express the difference between
the area radius (and in the axi-symmetric case also spin) associated
with two different MTSs $S_1$ and $S_2$ and flux of energy (and
angular momentum) across the portion $\Delta H$ of the DH bounded by
$S_1$ and $S_2$ \cite{ak,akrev}:
\be \frac{R_2-R_1}{2G} =  \int_{\Delta H} |d R|\, T_{ab}
\th^{\,a}\l^b\,d^3V  + \frac{1}{16\pi G}\,\int_{\Delta H} |d R|\, (
|\sl|^2 + 2|\zeta|^2)\,d^3V\, , \label{balanceR} \ee
where as before $\sl_{ab}$ is the shear of the outward pointing null
normal $\ell^a$ to the MTSs and $R$ is the area radius of the MTSs,
and where $|dR| = (q^{ab}\,D_aR D_b R)^{{1/2}}$ and the vector
field $\zeta^a$ tangential to each $S$ is defined by:
\be \label{zeta} \zeta^a := \t{q}^{\,ab}\, \rh^{\,c}\,\grad_c\l_b=
\t{\w}^a + \t{D}^a\ln |d R|\, . \ee
For the horizon spin, we have \cite{ak,akrev}
\be \label{balanceJ} J^\vp[S_2] - J^\vp[S_1] = -\int_{\Delta H}
\left( T_{ab}\th^{\, a}\varphi^b\,+\,\frac{1}{16\pi G} (K^{ab}- K
q^{ab})\, \Lie_\varphi {q}_{ab} \right)\, d^3V .\ee
These two balance laws follow directly from Einstein's equations. On
the conceptual side, they are significant because (unlike, say,
Hawking's area theorem for event horizons) they provide a detailed
link between the changes of physical quantities defined on $S_2$ and
$S_1$ and energy and angular momentum fluxes across the portion
$\Delta H$ bounded by them. In this respect, they are completely
analogous to the balance laws for the Bondi energy momentum and
angular momentum at null infinity. On the practical side, because
the quantities that appear in the integrand of the right and side
can be calculated independently of those that appear on the left
side of these equations, these balance laws can serve as internal
checks on accuracy of numerical simulations. We will now show that
there are balance laws associated with multipole moments \emph{that
share all these features.}

As in section \ref{s3.3}, let us begin with the spin multipoles.
Note first that, apart from overall constants that are needed for
dimensional reasons, the spin multipole moment $L_\phi [S]$ of Eq.
(\ref{L2}) is obtained simply by replacing the axial Killing vector
$\vp^a$ in the definition of the horizon spin \cite{ak,akrev},
\be J^\vp [S] := - \f{1}{8\pi G}\, \oint_S \t\omega_a\, \vp^a d^2V
\, ,\ee
with \emph{any} permissible divergence-free vector field $\phi^a$.
Therefore the balance law (\ref{balanceJ}) readily generalizes to:
\be \label{balL} L_{\phi} [S_2] - L_{\phi}[S_1]=-\int_{\Delta H}
\left( 4 \pi G T_{ab}\th^{\, a}{\phi}^b\, +\frac{1}{4}\, (K^{ab}- K
q^{ab})\, \Lie_{\phi} q_{ab} \right) \, d^3V. \ee
This generalization also has a direct analog at null infinity, where
one can introduce balance laws not just for the 4-momentum and
angular momentum but for charges associated with \emph{any} of the
generators of the infinite dimensional BMS Lie algebra
\cite{scri0,scri1,scri2,scri3}. Finally, we can also obtain a
differential balance law directly from the definition (\ref{L}) of
the spin multipole moments:
\be \frac{d L_{\phi}}{dv}  =  -\frac{1}{2} \oint_S \Lie_X
(\t\omega_c \phi^c \, \epsilon_{ab})\,  = -\frac{1}{2}\, \oint_S
\left[ \Lie_X(\t\omega_a) \phi^a  + 2 (\dot{R}/R) \t{\omega}_a\,
\phi^a \right]\, d^2V , \ee
where we have used the fact that $\phi^a$ is a permissible weighting
field. The structure of the right hand side of this equation is
quite analogous to that associated with the `BMS-fluxes' at null
infinity \cite{scri0,scri1,scri2,scri3}.

As one might expect from section \ref{s3.3}, the situation is the
same for the other set of moments, $I_\phi [S]$: One only has to
replace $\t\omega_a$ with $- C_a = -(\Gamma_a - \mathring\Gamma_a)$:
\be \frac{d I_{\phi}}{dv} = \frac{1}{2} \oint_S \left[ \Lie_X(C_a)
\phi^a  + 2 (\dot{R}/R) C_a\, \phi^a \right]\, d^2V. \ee

\section{Approach to an axi-symmetric isolated horizon} \label{s4}

In this section we will consider the physically interesting
situation in which a \emph{generic} DH settles down to an
axi-symmetric IH. Then, on the IH portion we can introduce a
convenient basis $\vp_{l,m}^a$ of divergence-free vectors using the
basis functions $Y_{l,m}$ made available by axi-symmetry. By
transporting them along the canonical vector field $X^a$ to the DH
portion we will obtain a convenient basis also on the DH portion,
thereby converting the multipoles $I_\phi, L_\phi$ defined in
section \ref{s3} to \emph{a set of numbers} $I_{l,m},\, L_{l,m}$
also on the DH. These can be readily evaluated in numerical
simulations of black hole formation to study the approach to
equilibrium. 

This section is of direct interest to numerical relativity because:
i) one expects the final IH in physical situations to be the Kerr IH
and therefore axi-symmetric; ii) these moments are better suited to
unravel universalities, if any, in the approach to equilibrium; and,
iii) In most circumstances it would suffice to track just the first
few multipoles. Therefore, for convenience of this readership, we
have attempted to make this section self-contained.

\subsection{The setting}
\label{s4.1}

Let us begin with a brief summary of the notation, collecting in one
place the terminology used to denote the numerous fields that
feature this analysis.
\begin{figure}
  \begin{center}
  \includegraphics[height=12cm]{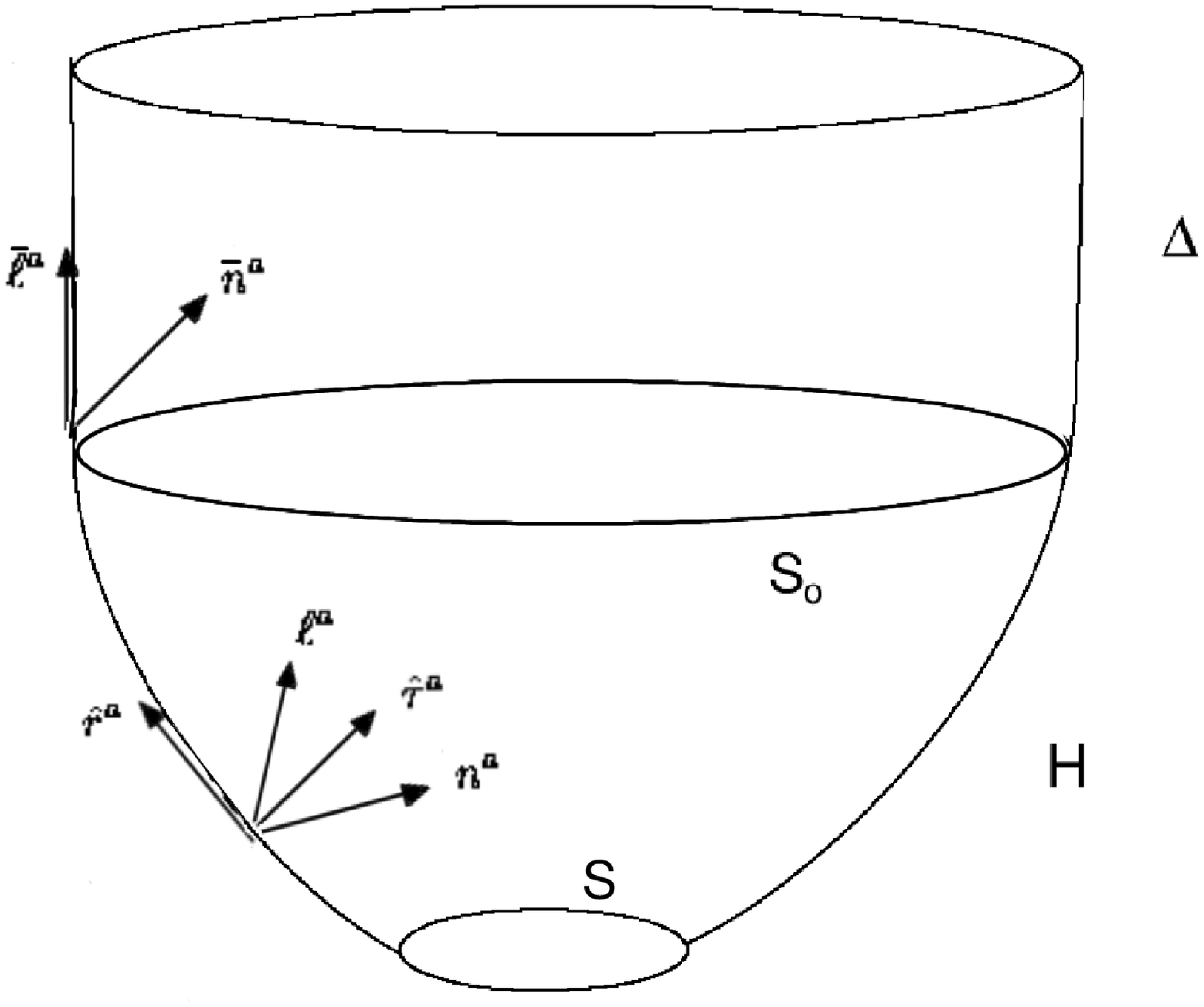}
  \caption{A quasi-local horizon $M$. The past portion of $M$ consists
  of a dynamical horizon $H$: this portion is space-like and foliated by
  marginally trapped surfaces $S$. $\th^a$ is the unit time-like normal to
  $H$ and $\rh^{\,a}$ the unit space-like normal within $H$ to the foliation.
  Although $H$ is space-like, motions along $\rh^{\,a}$ can be regarded as
  `time evolution' with respect to observers at infinity. $H$ joins on to an
  isolated horizon $\Delta$ in the future, representing the equilibrium state
  of the black hole. $\Delta$ is null, endowed with a preferred null normal
  $\bar{\ell}^a$. The transition from $H$ to $\Delta$ occurs at $S_o$.}
  \label{dhfig}
  \end{center}
  \end{figure}
Consider a quasi-local horizon $M$ with two parts: A DH $H$ in the
past that is joined on to an IH at a 2-sphere cross-section $S_o$
(see FIG. \ref{dhfig}). We will assume that $M$ is a 3-dimensional,
$C^{k+1}$ sub-manifold of space-time and the space-time metric
$g_{ab}$ is $C^k$ with $k \ge 2$. We will denote by $q_{ab}$ the
pull-back of $g_{ab}$ to $M$; thus $q_{ab}$ has signature +,+,+ on
the portion $H$ of $M$ and 0,+,+ on the portion $\Delta$. The
space-time connection $\nabla_a$ induces a natural connection on $M$
which we denote by $D_a$. It satisfies $D_a q_{bc} =0$ on all of
$M$.

The (future pointing) null normal to the IH $\Delta$ will be denoted
by $\lb^a$. The second (also future pointing) null normal to $S_o$
will be denoted by $\nb^a$. As noted below, these null vector fields
admit natural smooth extensions to $H$ and everywhere on $M$ we
choose them to satisfy the normalization $\lb^a\, \nb_a = -2$.
Finally, we define a 1-form $\omega_a$ on all of $M$ via
$t^a\omega_a = - (1/2) \nb_b t^a\,{\nabla_a}\, \lb^b$ for all $t^a$
tangential to $M$. $\omega_a$ represents the rotational 1-form on
$\Delta$ while, as discussed below, on $H$ it equals the
$\t\omega_a$ defined in (\ref{mm2}) modulo a gradient which drops
out of the expression of multipole moments. We will denote the axial
symmetry vector field on the IH by $\vp^a$. But we \emph{do not}
assume that the DH is axi-symmetric. It is allowed to have arbitrary
distortions in its intrinsic and extrinsic geometry.

On the dynamical horizon $H$, we will denote the unit normal to $H$
in the space-time manifold $(\man, g_{ab})$ by $\th^{\,a}$ and the
unit normal to the MTSs $S$ within $H$ by $\rh^{\,a}$. Then $\ell^a
= \th^{\,a} + \rh^{\,a}$ and $n^a = \th^{\,a} - \rh^{\,a}$ are the
two null normals to the MTSs, with $\ell^a n_a = -2$. Let the MTSs
be the level surfaces of a $C^{k+1}$ function $v$. We will assume
that $S_o$ is the uniform limit of MTSs and is thus labeled by $v=
v_o$. One can continue the foliation in the future on the IH such
that $v$ is the affine parameter of the null normal $\lb^a$. We will
do so. On $H$ we set $2b = |dv|^{-1} = (q^{ab}D_a v D_b v)^{-1/2}$.
The area-radius of the horizon cross-sections will be denoted by
$R$; $R$ increases monotonically with $v$ and remains constant on
$\Delta$. The extrinsic curvature of $H$ within space-time $(\man,
g_{ab})$ is denoted by $K_{ab}$. The intrinsic 2-metric on the MTSs
$S$ is denoted by $\t{q}_{ab}$ and its derivative operator by
$\t{D}_a$. More generally, if there is an ambiguity in the notation,
we use a tilde to denote fields that are intrinsic to the MTSs.\\

Since $M$ is space-like in the past and null in the future, the
transition at $S_o$ is somewhat subtle. Let us collect the basic
facts (from \cite{ak,bf} and the Appendix) that are needed in the
analysis of multipoles.
\itemize

\item i) $b^2$ admits a $C^k$ limit to $S_o$ and
    vanishes there.
\item ii) On $H$ we have $\lb^a = b\ell^a$ and $\nb^a = b^{-1} n^b$.
    Thus, while $\ell^a$ is well-defined on $H$ it diverges at $S_o$
    and cannot be extended to $\Delta$. $(\lb^a, \nb^a)$, on the other
    hand, are smooth on all of $M$.
\item iii) The vector field $V^a:= 2b\, \rh^{\, a}$ on $H$ admits a
    smooth extension to $\Delta$ and equals $\lb^a$ on $\Delta$. On
    all of $M$, $V^a$ can be regarded as an evolution vector field
    with zero shift. Indeed, since $V^aD_a v =1$ everywhere on $M$,
    $v$ serves as the affine parameter of $V^a$. Finally, $b^2 =
    V^a\, \bar\ell^b g_{ab}$.
\item iv) If we set $\dot{R} = dR/dv$, then both $\dot{R}$ and $b$
    are non-zero on $H$ but vanish on $S_o$ and remain zero on
    $\Delta$. The field $b_o^2 : = b^2/\dot{R}$ is non-zero and smooth
    on $S_o$.
\item v) the rotational 1-form $\omega_a$ which is well-defined
    everywhere on $M$ and the 1-form $\t\omega_a := \t{q}_a{}^b\,
    K_{bc}\, \rh^{\,c}$ defined on $H$ are related by $\omega_a =
    \t{\omega}_a - \t{D}_a \ln b_o$.

\subsection{Steps for numerical simulations}
\label{s4.2}

The general multipole moments defined in \ref{s3} are somewhat
abstract: Given any MTS $S$, the $I_\phi [S],\, L_\phi [S]$ can be
regarded as linear mappings from permissible divergence-free vector
fields $\phi^a$ on $H$ to real numbers. As noted in the beginning of
this section, in physical situations we expect $H$ to join on to an
\emph{axi-symmetric} isolated horizon in the future, in fact the
Kerr IH. We can exploit this extra structure by first locating a
preferred basis $\vp_{l,m}^a$ of divergence-free $\phi^a$ on
$\Delta$ and then drag it along the preferred dynamical vector field
$X^a$ (of (\ref{X}), spelled out again below) to the DH portion $H$
of $M$. Put differently, we can now define the weighting functions
$Y_{l,m}$ on the axi-symmetric IH and drag them down to $H$ along
$X^a$, making them explicitly time independent. Given this basis of
weighting functions, one can now replace the multipole moments
$I_\phi [S]$ and $L_{\phi}[S]$ on $H$ with just a set of numbers
$I_{l,m}[S],\, L_{l,m}[S]$ which are well-suited to study, in an
invariant manner how the black hole reaches its equilibrium in any
one numerical simulation. Furthermore, now one can also compare the
results of two \emph{different} simulations since one just has to
compare numbers $I_{l,m}[S],\, L_{l,m}[S]$ associated with the MTSs
$S$ with the same area. In practice the first few moments are likely
to contain the most interesting information on passage to
equilibrium.

Consider, then, a numerical simulations of a black hole formation.
The world tube of MTSs found after a common horizon forms provides
us with the 3-manifold $M$ of FIG. \ref{dhfig}. To extract multipole
moments, one has to carry out the following steps.

i) In the portion $H$ of $M$ on which the area of the MTSs increases
monotonically, calculate the following quantities: a) The 3-metric
$q_{ab}$, b) the intrinsic 2-metric $\t{q}_{ab}$, c) the area radius
$R$ of each MTS $S$, so that the area of $S$ is $4\pi R^2$, d) the
unit normal $\rh^{\,a}$ to each $S$, and, e) the trace of the
extrinsic curvature $\t{K}$ of each $S$ within $H$.%
\footnote{In numerical simulations, one solves the initial value
problem using a 1-parameter family of Cauchy surfaces $\Sigma_t$ and
locates the outermost marginally trapped surface $S_t$ on each
$\Sigma_t$. The DH $H$ is the world tube of these 2-surfaces.
Therefore, fields which are naturally available refer to $\Sigma_t$
and $S_t$ and some extra steps are necessary to extract the fields
such as $q_{ab}$ and $\rh^{\,a}$ we need here. These are described
in section III of \cite{skb}; see in particular Eqs (3.3)-(3.5),
(3.9) and (3.13) in that section.}

ii) Find the 1-form $\t{\omega}_a := \t{q}_a{}^b K_{bc}\, \rh^{\,c}$
on each MTS S. This is the `seed' that will generate the (geometric)
spin moments $L_{l,m}$. Find the scalar curvature $\R$ of the metric
$\t{q}_{ab}$ on each MTS $S$, which will serve as the seed for the
(geometric) mass moments $I_{l,m}$. Taking the required second
derivatives may introduce undesirable numerical errors (see,
however, \cite{kvf3}). If so, it may be more convenient to introduce
a complex orthonormal dyad $m^a, \bar{m}^a$ on each $S$ and
calculate the so-called `spin connection' $\Gamma_a$ via $\t{D}_a
m_b=: i\Gamma_a m_b$. This 1-form $\Gamma_a$ can also serve as the
seed to calculate the second set of moments $I_{l,m}$.

iii) Now introduce a coordinate $v$ on $M$ such that the MTSs are
the $v= {\rm const}$ surfaces and the vector field $V^a = |dv|^{-1}
\rh^{\,a} \equiv 2b \rh^{\,a}$ smoothly becomes the null normal
$\lb^a$ to the IH in the future region of $M$.

iv) In the next step, construct the dynamical vector field $X^a$
(that will be used to transport the weighting functions $Y_{l,m}$
from the IH to the DH). On each MTS $S$, find the function $g$ via
(\ref{g})
\be  \t{q}^{ab} \t{D}_a \t{D}_b g = -(2b \t{K} - 2{\dot{R}}/{R})
\quad\quad {\rm and} \quad\quad \oint_S \, g\, d^2 V = 0\ee
and define the `shift' field $N^a$ via $N^a = \t{q}^{ab} \t{D}_b g$.
Then the dynamical vector field $X^a$ is given by $X^a = V^a + N^a$.
(As we approach the IH, $g$ and $N^a$ tend to zero and $X^a$ joins
on smoothly with $\lb^a$.)

v) On the 2-surface $S_o$ where the DH $H$ joins on to the IH
$\Delta$ (or anywhere to its future), find the axial Killing field
$\vp^a$, e.g., using the algorithm described in
\cite{dkss,kvf1,kvf2,kvf3,kvf4}. In practice, one would expect the
geometry to become axi-symmetric within numerical errors already at
late times on the DH and one can then find $\vp^a$ on that MTS
without having to locate $S_o$. On the MTS $S$ on which $\vp^a$ is
found, by the standard procedure developed in \cite{skb} using
\cite{aepv}, find the basis functions $Y_{l,m}$ (defined by the
canonical `round' metric determined by $\vp^a$ and the $\t{q}_{ab}$
on that MTS).

vi) Drag these weighting functions to any MTS $S$ of interest via
$\mathcal{L}_X Y_{lm} =0$. Construct the multipole moments $L_{l,m}$
on that $S$ using (\ref{L}):
\be L_{l,m} = - {\textstyle \f{1}{2}}\, \oint_S (\epsilon^{ab}\, \t{D}_b
Y_{l,m})\, \t{\omega}_a\, d^2 V = - {\textstyle \f{1}{2}}\, \oint_S
(\epsilon^{ab}\, \t{D}_b Y_{l,m})\, {\omega}_a\, d^2 V\, . \ee
Thus, one has to either evaluate the 1-form $\t\omega_a$ that refers
to the extrinsic curvature $K_{ab}$ of $H$ in the 4-dimensional
space-time, \emph{or} the rotational 1-form $\omega_a$ that refers
to $\lb^a$ and $\nb^a$, whichever is numerically easier. Next
consider the moments $I_{l,m}$. For $l =0$, we have $I_{0,0} =
2\pi$, a topological invariant. For $l \not=0$, we again have two
avenues, given in the following two equivalent definitions:
\be I_{l, m} := {\textstyle \f{1}{4}}\, \oint_S \R\, Y_{l,m}\, d^2V
= {\textstyle \f{1}{2}} \oint_S (\epsilon^{ab}\, \t{D}_b Y_{l,m})\,
  (\Gamma_a - \mathring\Gamma_a)\, d^2V \ee
where $\mathring\Gamma_a$ is a fiducial connection; we can set
$\mathring\Gamma_a = -(1/R^2)\, \cos\theta\, \partial_a \phi$ in the
coordinates used to express the $Y_{l,m}$. The second form may be
more helpful if there are large numerical errors in computing the
scalar curvature $\R$. Finally, the mass and spin multipoles
$M_{l,m}$ and $J_{l,m}$ can be constructed by multiplying these
geometric multipoles with appropriate dimensionful factors
\cite{aepv,akrev,skb}.

This six step procedure enables one to compute the geometric
multipole moments and study their evolution during the highly
dynamical phase immediately after the formation of the common
horizon. Computing these moments in examples is likely to bring out
patterns in the way black holes shed their hair and approach the
final equilibrium state, which in turn may enable one to uncover any
universalities this process may have. In particular, on each MTS $S$
the procedure provides a \emph{spin vector} since generically
$L_{1,m}$ will be non-zero even when $m\not=0$. The `direction' of
the spin vector can change during the dynamical phase and the black
hole would shed the $x$ and the $y$ components of this spin vector
entirely as it reaches equilibrium. Does this process simply vary
from case to case, depending strongly on the structure of the common
horizon at its birth, or is there some underlying law that relates
it to, say, the angular momentum radiated away to null infinity?

Note that all the moments are anchored in the structure provided by
the final equilibrium state of the black hole. The change in the
mass dipole, for example, tells us how the black hole loses its
3-momentum \emph{with respect to its final equilibrium state}. In
fact a natural `home' for the multipoles is provided by the tangent
space at the point $i^+$ at future time-like infinity: The $I_{l,m}$
(or $L_{l,m}$), for example, can be naturally regarded as
constituting an $l$th rank, trace-free, symmetric tensor in the
tangent of $i^+$, all of whose indices are orthogonal to the final
Bondi 4-momentum of the black hole.

Finally, as discussed in section \ref{s3.4}, there are balance laws
that bring out the fact that the multipoles evolve in time in
response to fluxes of physical fields across the DH $H$. In section
\ref{s3.4} we considered the multipole moments weighted by
permissible divergence-free vectors $\phi^a$. We now have a
preferred basis $\phi^a_{l,m}$ constructed from spherical harmonics
$Y_{l,m}$, given in Eq. (\ref{amm}). Therefore we can rewrite the
balance laws using the $Y_{l,m}$ as weighting fields. Given two MTSs
$S_1$ and $S_2$, the difference between the spin multipoles
associated with them can be expressed in terms of a flux across the
portion $\Delta H$ of $H$, bounded by $S_1$ and $S_2$:
\be \label{balLlm} L_{l,m} [S_2] - L_{l,m}[S_1] = -\int_{\Delta H}
\! \left(4 \pi G T_{ab}\th^{\, a}\, \epsilon^{bc}\, D_c Y_{l,m}\,
+\frac{1}{2} (K^{ab}- K q^{ab})\, D_a (\epsilon_{bc}\, D^c Y_{l,m})
\right) \, d^3V\, .\ee
Similarly, for the $I_{l,m}$, we have the balance law:
\ba \label{balanceIlm} I_{l,m} [S_2] - I_{l,m} [S_1] &=&
\int_{\Delta H} \frac{|dR|}{2 R} \left(|\sl|^2 + 2
|\zeta|^2 + 16 \pi G T_{ab} \th^{\,a}
\l^b \right) Y_{l,m} d^3V  \\
&+&\int_{\Delta H} |dR| \left(\frac{1}{4}Y_{l,m} \; \partial_R
\R+\frac{1}{R}\zeta^a \partial_a\, Y_{l,m} \right) d^3 V\, . \ea
On the IH, the flux integral on the right vanishes identically and
the multipoles are conserved. On the DH portion, on the other hand,
these balance laws could provide useful checks for numerics since
the left and right sides refer to entirely different fields and they
are equal only when Einstein's equations are satisfied.

\emph{Remarks:}

1. Throughout this analysis we have restricted ourselves to the
dynamical horizon $H$ of the final black hole. Suppose we begin with two
widely separated black holes which coalesce. Before the merger, we
would have two distinct DHs, say $H_1$ and $H_2$. In the distant
\emph{past} these would join on to two distinct IHs $\Delta_1$ and
$\Delta_2$, each of which would be well-modeled by a Kerr IH and
hence axi-symmetric. Therefore, using the procedure described in this
section, on \emph{each} of these two quasi-local horizons, one would
be able to define multipole moments $I_{l,m}$ and $L_{l,m}$
separately, where the required weighting functions $Y_{l,m}$ would
now be transported from $\Delta_1$ to $H_1$ and from $\Delta_2$ to
$H_2$. In particular, one would be able to study the evolution of
the spin $L_{1,m}$ of each individual black hole. However, at
present the DH framework cannot describe the merger phenomenon
simply because $H_1\cup H_2 \cup H$ is not a DH. Therefore, there is
no simple relation between the two sets of multipoles prior to the
merger and the set of multipoles after the merger.

2. Nonetheless, using the structure available at null infinity one
can discuss \emph{global} balance laws. Recall first that the total
Arnowitt-Deser-Misner energy-momentum is well-defined at the point
$i^o$ at spatial infinity \cite{ah}, or, equivalently, in the
distant past of $\mathcal{I}^+$ \cite{am}. Denote it by $P^a_{\rm
initial}$. (Note that $P^a_{\rm initial} \not= P^a_1 + P^a_2$ in
general, e.g., because of the potential energy in the system.)
Similarly, in the distant future, the mass monopole of the IH
determines the final Bondi 4-momentum $P^a_{\rm final}$. Both can be
thought of as living in the 4-dimensional vector space dual to the
space of BMS translations. Therefore, it is meaningful to consider
their difference $P^a_{initial} - P^a_{final}$ and this is precisely
the Bondi 4-momentum radiated across $\mathcal{I}^+$ in the
dynamical coalescence for which we have an independent formula
\cite{am}.

The situation with angular momentum is similar but more subtle. The
total (Lorentz) angular momentum of the system $M^{ab}_{\rm
initial}$ is a well-defined mapping \cite{scri2,scri3} from the
Lorentz Lie algebra of the BMS Lie algebra, picked out by the fact
that the Bondi news goes to zero as one approaches $i^o$
\cite{np,aa}. Again, $M^{ab}_{\rm initial}$ is not simply related to
$S_1^a + S_2^a$, e.g., because it also contains a contribution due
to the orbital motion. The final angular momentum, $M^{ab}_{\rm
final}$, on the other hand \emph{is} determined entirely by the
final spin of $H$ because in the distant future we only have a
single black hole. However, it refers to a distinct Lorentz sub-Lie
algebra of the BMS Lie algebra now selected by the fact that the
Bondi news goes to zero in the distant future. (The two Lorentz
sub-algebras agree only in the special circumstance in which the
integral of the Bondi news along every generator of $\mathcal{I}^+$
vanishes \cite{np,aa}.) Therefore it is not meaningful to take the
difference $M^{ab}_{\rm initial} - M^{ab}_{\rm final}$. Rather, in
place of $M^{ab}_{\rm initial}$, we have to consider the angular
momentum $\bar{M}^{ab}_{\rm initial}$, again evaluated in the
distant past of $\mathcal{I}^+$ but associated with the Lorentz
sub-group picked out by the Kerr geometry in the distant future.
This $\bar{M}^{ab}_{\rm initial}$ is well-defined but \emph{not} the
same as $M^{ab}_{\rm initial}$ even conceptually. The difference
$\bar{M}^{ab}_{\rm initial} - M^{ab}_{\rm final}$ is well-defined
because both quantities now refer to the \emph{same} Lorentz
sub-group of the BMS group. Furthermore, by the balance laws
\cite{scri2,scri3}, this is precisely the angular momentum
(associated with the common Lorentz group) radiated across scri.

To summarize, the balance laws are meaningful both for the
4-momentum and angular momentum, although in the case of angular
momentum, to compare `apples with apples', we have to drag the
weight functions corresponding to the canonical Lorentz group in the
distant future of $\mathcal{I}^+$ to distant past. Thus, there is no
simple relation between the initial spins $S_1$ and $S_2$ of the
individual black holes, the final spin $S$ of the common black hole
and the angular momentum radiated away across $\mathcal{I}^+$. For
the 4-momentum, we do have a balance law relating $P^a_{\rm
final},\,\, P^a_{\rm initial}$ and the 4-momentum radiated away
across $\mathcal{I}^+$. However, unless $P^a_{\rm initial} \approx
P^a_1 + P^a_2$, there is no simple relation between the initial
4-momenta of individual black holes and the 4-momentum of the
single, final black hole.

\subsection{Comparisons} \label{s4.3}

We will conclude section \ref{s4} with a discussion of the relation
of this construction with similar ideas in the literature.

As we showed in section \ref{s3}, ours is a genuine generalization
of the definition \cite{skb} used in cases when the DH is
axi-symmetric. The generalization is both technically non-trivial
and conceptually important because in the early stage of the post-merger phase,
the DH is generally very far from being axi-symmetric. We allowed
the DH to be generic and assumed axi-symmetry only for the IH
representing the final equilibrium. Nonetheless, if the entire
quasi-local horizon is axi-symmetric as in \cite{skb}, then on any
MTS $S$ our weighting functions $Y_{l,m}$ coincide with those
determined intrinsically on $S$ using the restriction of the axial
symmetry $\vp^a$ to $S$.

There is another generalization in the literature, due to Owen
\cite{ro}. That definition has the non-trivial feature that, while
it uses only the DH portion of $M$ without reference the final IH as
in section \ref{s3}, the multipoles are a set of numbers as in
section \ref{s4.2}. This is achieved using a construction that is
local to each MTS $S$ of the DH. In particular, the weighting
functions used in \cite{ro} are eigenfunctions of certain elliptic
operators constructed entirely from the geometry of the MTS; one
does not transport them from a final axi-symmetric state. For the
mass moments, the elliptic operator is just the intrinsic Laplacian
(determined by the physical metric $\t{q}_{ab}$) but for the spin
moments a different, 4th order elliptic operator is used to ensure
that, if the DH \emph{is} axi-symmetric, the general procedure
provides the well-established spin vector. These multipoles are
distinct from ours and, generically, in the axi-symmetric case they
are different also from the multipoles introduced in \cite{skb}.

The main differences from our definition are the following. First,
while we use the same weighting functions for both sets of
multipoles, Owen used different weighting functions. The second and
more important difference is that we transport the weighting
functions by dragging them from the final equilibrium configuration
so that they are constant along the dynamical vector field $X^a$. By
contrast, Owen's weighting functions are determined by the local,
time varying geometry. Owen's construction has the advantage of
being `local in time', i.e., being covariant with respect to the
geometry of each individual MTS. Our procedure is covariant only
with respect to the geometry of the quasi-local horizon $M$ as a
whole. On the other hand, because our weighting functions on any MTS
are `the same' as those in the final equilibrium state, our
multipoles directly capture the dynamics of the horizon geometry
encoded in $\R$ and $d\t\omega$ in the passage to equilibrium; one
compares `apples with apples'.

To clarify this issue of time dependence, it is useful to recall the
conceptual parallel between the definition of multipoles on a DH and
that of the `BMS charges' at null infinity
\cite{scri0,scri1,scri2,scri3} we used in Remark 2 at the end of
section \ref{s4.2}. The BMS charges are integrals over 2-sphere
cross sections of null infinity of `seed' physical fields, weighted
by functions that refer to the BMS symmetry corresponding to the
charge. (In this analogy, the cross sections of null infinity play
the role of the MTS $S$ on the DH, the `seed' physical fields
correspond to our $\R, d\t{\omega}$ on $S$, and the weighting
functions, to the $Y_{l,m}$ used here.) In the BMS case, given any
cross section of null infinity, using its intrinsic geometry, one
can find weighting functions corresponding to a specific Lorentz
sub-Lie algebra of the BMS Lie algebra and construct six charges
that represent the Lorenz-angular momentum at (the retarded instant
of time represented by) that cross-section. However, generically,
different cross-sections select \emph{different} Lorentz sub-Lie
algebras of the BMS Lie algebra and therefore it is not meaningful
to compare the resulting Lorentz charges on one cross section to
that on another. To compare `apples with apples', one has to use the
\emph{same} Lorentz sub-group of the BMS group. This is achieved by
appropriately \emph{transporting} the generators (or weighting
functions) corresponding to the Lorentz subgroup used on the first
cross-section to the second cross-section and carrying out the
2-sphere integral with these \emph{transported} generators which, in
general, are distinct from those determined intrinsically by that
cross-section. Thus, the notion of the `same' Lorentz sub-Lie
algebra refers to the structure of the 3-dimensional null infinity
as a whole; it cannot be captured by working locally on each
cross-section. And it is only when the `same' Lorentz generators are
used that the change between the two sets of Lorentz charges refers
to the change in the same \emph{physical} quantities. There is no
`contamination' due to a change in the weighting function itself,
which would have occurred if we had used the generators selected by
each cross section separately.

On quasi-local horizons, our procedure embodies this spirit in that
our transport of weighting fields $Y_{l,m}$ from the final isolated
horizons $\Delta$ to the dynamical horizon $H$ is analogous to the
transport of the Lorentz generators which is necessary for
comparisons. Therefore, our multipoles $I_{l,m} [S],\, L_{l,m} [S]$
on any MTS $S$ of the DH can be meaningfully compared to those in
the final equilibrium state. They are thus well-adapted to meet the
goal of this paper: capturing the physics of dynamics that makes the
black hole shed its `hair' in its \emph{approach to} equilibrium.
Owen's goal was different. The focus there was to investigate the
structure of the final state itself and the analysis provided
evidence that it is Kerr. To meet that goal, it is not necessary to
transport the weighting fields.

Finally, over the last two years there has been notable interest in
numerical simulations whose goal is to visualize the strong field
regime around black holes in terms of the so-called `tendex and
vortex lines' \cite{vt1,vt2}. The idea is to repeat the strategy
that has been so successful in electrodynamics where pictorial
representations of the magnetic lines of force often provide good
intuition for the complicated dynamics, e.g., in problems involving
neutron stars. In the case of black holes, the gravitational lines
of force are obtained using the eigen-directions of the electric and
magnetic parts of the Weyl tensor,
\be E_{ab} =  C_{a c b d} \tt^c \tt^d\, , \quad {\rm and} \quad
B_{ab} = {}^\star C_{a c b d} \tt^c \tt^d
={\textstyle{\f{1}{2}}}\epsilon_{ac}^{\phantom{ac}pq}C_{pq b d}
\tt^c \tt^d , \ee
with respect to a space-time foliation to which $\tt^a$ is the unit
time-like normal field. In the Kerr space-time, one can use natural
foliations, the lines cross the MTSs, and their visual properties
provide intuition for physical effects of the near horizon, strong
gravitational field. These images are also useful when one considers
perturbations around Kerr. However, in a truly non-linear, dynamical
situation, e.g., at the formation of the common horizon during
generic black hole collisions, there are no natural space-time
foliations. Since the lines of force are tied to foliation choices
that are made by extrapolating one's intuition based on the
stationary Kerr geometry (and perturbative dynamics thereon) these
visual images cannot be used to draw reliable conclusions about the
\emph{physics} of dynamical processes in the strong field,
near-horizon geometry. Multipole moments introduced in this paper
serve a complementary role. In particular, it would be instructive
to develop programs to visualize the distortions in the geometry and
the angular momentum content of the dynamical horizon membrane. In
Kerr space-times, the geometrical intuition provided by these
visualizations would not be as rich as that provided by vortexes and
tendexes where the lines of force extend beyond the horizon, all the
way to infinity. But in the strong field and highly dynamical
regime, the intuition these multipoles provide would capture a more
accurate depiction of the actual, invariant physics.


\section{Discussion}
\label{s5}

There is growing evidence that, in general relativity, the final
equilibrium state of black hole horizons is extremely
well-approximated by the Kerr horizon. However, immediately after
its formation, the common horizon that surrounds all matter and
individual black holes is highly dynamical and its evolution varies
from case to case. In this paper we have introduced multipole
moments to gain physical insights into the strong field dynamical
processes that efficiently smoothen all the distortions, leading to
an universal final geometry.

We presented two sets of ideas. The first, discussed in section
\ref{s3}, is most useful on DHs $H$. It associates with each MTS $S$
of $H$ multipole moments $I_\phi [S], \, L_\phi[S]$, where the
weighting functions $\phi$ are a set of `time independent' vector
fields $\phi^a$ which are tangential to each $S$ and divergence-free
on them. On any given DH, the evolution of these multipoles captures
the dynamics of the transition to equilibrium in a coordinate and
slicing independent fashion. The second idea, presented in section
\ref{s4}, is applicable only in a setting in which the DH is joined
on to an \emph{axi-symmetric} IH in the future. However, from
physical considerations, this is not a genuine restriction because,
as we just noted, one expects the final equilibrium state to be the
Kerr IH. In this case, one can introduce a convenient basis
$\vp^a_{l,m}$ in the space of weighting fields $\phi^a$, labeled by
spherical harmonics $Y_{l,m}$ that are determined on $H$ in an
invariant fashion by the future axi-symmetric structure.
Consequently, now the multipole moments on any MTS $S$ are
\emph{just a set of numbers} $I_{l,m}[S], L_{l,m}[S]$. As we saw in
section \ref{s4.2}, their definitions are well-suited for numerical
simulations. Not only can one use them to monitor dynamics on any
one DH, but they also enable one to \emph{compare} results of
distinct simulations. (This is not possible with $I_\phi [S],\,
L_\phi [S]$ because one does not have a canonical identification
between the divergence-free vector fields $\phi^a$ on the two DHS
obtained in two distinct simulations.) Also, these multipoles
provide tools to physically interpret the dynamical process. For
example, $L_{1,m}[S]$ provides a well-defined notion of the
\emph{spin-vector} during the dynamical phase. Tracking the
evolution of the direction of the spin is likely to provide new
insights. More generally, by explicitly evaluating a few low $l$
multipoles and monitoring their evolution, numerical simulations
should be able to find any patterns or universalities in the manner
black holes shed their hair. We also provided formulas for fluxes of
these multipoles. Since they are strict consequences of field
equations, they can serve as analytic checks on numerics in the
strong field and highly dynamical regimes.

This framework is well-suited to analyze a number of issues. Recall
first that, over the years, perturbative investigations have
provided strong indications that the passage to equilibrium may have
some universal features. In particular, the Price's law and
increasing evidence in its favor \cite{price1,price2,rd}, the
success of the close limit approximation of Price and Pullin
\cite{pp,gnpp}, and the universality of quasi-normal ringing
\cite{price2} all suggest that, although the strong field dynamics
after the formation of a common horizon is highly non-linear, it has
a deep underlying simplicity. However, to date the investigations in
the strong field limit have been restricted to spherical symmetry
\cite{rd}. In this case, there is no gravitational radiation, the DH
has no hair, and its dynamics is rather simple and fully understood
\cite{bi}. The central questions concerning the dynamical processes
that wash away the distortions and non-trivial angular momentum
structure of the DH simply do not arise. Therefore, numerical
studies of the time evolution of multipoles $I_{l,m}[S],\, L_{l,m}
[S]$ in the general case, far removed from spherical symmetry, could
lead to fresh and interesting insights. As the DH reaches its
equilibrium, is there a correlation between rate at which it sheds
its multipoles and, say, Price's law? For example, recent numerical
simulations suggest that the end point of the collision of two
spinning black holes can be a Schwarzschild black hole \cite{jh}. In
this case the DH would have to lose \emph{all} its multipoles except
the mass monopole. Is there a pattern to how they are lost? Is it
the case, as one would intuitively expect, that the high
$l$-multipoles die quickly while the low $l$ are dissipated more
slowly? Is there in fact a \emph{quantitative}, universal behavior?
Another example is provided by the `anti-kick' that is associated
with the post-merger phase of dynamics of binary black holes
\cite{jaramillo1}. There are general arguments to suggest that it
should be possible to account for this phenomenon in terms of the
behavior of the mass monopole and dipole of the DH that forms after
coalescence. Again, calculation of these moments and investigating
their dynamics are likely to provide new physical insights.

The physical process involved in the manner equilibrium is reached
is not directly intuitive because the DH lies \emph{inside} the
event horizon. Consequently, it does \emph{not} radiate away its
multipoles to infinity. Rather, distortions in the geometry and the
angular momentum structure of the DH are washed out by the radiation
that falls \emph{into} the black hole. But it appears that there is
a \emph{correlation} between what falls \emph{into} the DH and what
gets radiated away to null infinity. The qualitative picture is that
there is some radiation in the potential just outside the event
horizon, some of which falls into the black hole and the rest
escapes to infinity `remembering' the way it was correlated. At
first this scenario can seem rather far-fetched because it is
difficult to imagine processes responsible for this memory
retention. But the paradigm is supported by several recent
simulations \cite{eloisa,jaramillo1,jaramillo2,lr}. Multipole
moments defined here should help further develop these ideas.


\textbf{Acknowledgements:} We would like to thank Chris Beetle,
Jonathan Engle, James Healy, Jose Jaramillo, Pablo Laguna, Rob Owen, Georgois Pappas
and Kip Thorne for discussions. This work was supported in part by
the NSF grant PHY-1205388, the Eberly research funds of Penn State
and a Frymoyer Fellowship to MC.

\begin{appendix}
\section{Limiting behavior of physical fields}
\label{a1}

In this Appendix we sketch the limiting behavior of various fields
on the DH $H$, as one approaches the transition 2-surface $S_o$ that
joins $H$ with a non-extremal IH $\Delta$. These limits were used in
sections \ref{s3} and \ref{s4}. They also provide guidance for
numerical simulations in that they separate fields which are likely
to be easier to evaluate numerically from those that would be
challenging because they involve ratios of quantities, both of which
vanish or diverge in the limit. Finally, this discussion of the
limiting behavior should be helpful for further analytical work on
the approach to equilibrium.

Our notation is the same as in the main paper; see, e.g., section
\ref{s4.1}.

\subsection{The intrinsic and the extrinsic geometry of the DH}
\label{a1.1}

Since the DH $H$ is foliated by MTSs $S$, it is natural to decompose
the intrinsic metric $q_{ab}$ and the extrinsic curvature of $H$ as
follows:
\ba
q_{ab} & = & \t{q}_{ab}+ \rh_{\,a} \rh_{\,b}, \label{q}\\
K_{ab} &=& A \t{q}_{ab} + S_{ab} + 2\t\omega_{(a}\rh_{\, b)} + B
\rh_{\,a} \rh_{\,b} , \label{K}\ea
where, as in the main text, $\rh_{\,a}$ is the unit normal to the
MTSs $S$,\, $\t{q}_{ab}$ the intrinsic 2-metric on each $S$,\,
$S_{ab}$ is a symmetric trace-free tensor field on $S$ and
$\t{\omega}_a$ a 1-form field on $S$. We will investigate the
limiting behavior of these fields as we approach the limiting MTS
$S_o$ that joins $H$ to an IH $\Delta$.

As in the main text, let us introduce a `time' coordinate $v$ on the
entire quasi-local horizon $M$ such that the MTSs are the level
surfaces of $v$ and $v= v_o$ on $S_o$. Thus the portion $v<v_o$ on
$M$ corresponds to the DH and the portion $v>v_o$ corresponds to the
IH. We are interested in the behavior of various geometric fields as
$v$ approaches $v_o$ from below. We will assume that on the IH
portion of $M$, $v$ is the affine parameter of the null normal field
$\bar{\ell}^a$, i.e. $\bar\ell^a \partial_a\, v =1$. Given such a
function $v$, there is a unique vector field $V^a$ on $M$ such that:
i) on the DH, $V^a$ is normal to each MTS $S$, and, ii) satisfies
$V^a \partial_a v =1$ on all of $M$. Thus, $V^a$ is a smooth
extension of $\bar\ell^a$ on $\Delta$ to all of $M$. On $H$, $V^a$
is proportional to $\rh^{\,a}$:
\be \label{V} V^a =|dv|^{-1} \rh^{\,a}=:2 b\, \rh^{\,a},\quad {\hbox
{\rm with}} \quad  2b = \dot{R}\, |dR|^{-1}  \ee
where `dot' will denote the derivative with respect to $v$. It then
follows that  $V \cdot V= 4\, b^2$. Since $V^a$ is smooth and
coincides with $\bar\ell^a$ on $\Delta$, we conclude that $b^2$ is
smooth, vanishes at $v= v_o$, and remains zero for $v > v_o$.

Since the function $b$ features in the relation between the natural
null normals $\ell^a, n^a$ adapted to $H$ and the natural null
normals ${\lb}^a, {\nb}^a$ adapted to $\Delta$, its limiting
behavior dictates that of several fields. Let us therefore make a
small detour to specify the `rate' at which $b$ vanishes as we
approach $v=v_o$ from below. Note first that the rate of change of
area $A_S$ of a MTS $S$ on $H$ can be expressed as: $\dot{A}_S=
\oint_{S} {\cal L}_{V}\,( \epsilon_{ab} )$. Using the identity
${\cal L}_{V}\, \epsilon_{ab} = -\, b^2 \tnb \epsilon_{ab} $, and
expressing $\dot{A}_S$ in terms of the rate of change $\dot{R}$ of
the area-radius, we obtain $8 \pi R \dot{R}= -\oint_{S_v} b^2 \tnb
d^2V$. Therefore,
\be \lim_{v \to v_o} \oint_{S_v} \frac{b^2}{\dot{R}}\, \tnb\, d^2V=-
8 \pi R_o , \label{limitb0} \ee
where $R_o$ is the area-radius of $S_o$. Now, the integrand in
(\ref{limitb0}) is strictly negative for $v < v_o$ and $\tnb$ has a
well-defined limit $\tnb^{(o)}$ on $S_o$. Let us assume that we are
in a generic case and the limit is non-zero (a condition satisfied
on the
Kerr isolated horizon). 
Then it follows, e.g.  by Taylor expansion of fields in $v$, that
\be \label{b0} b_o= {b} (\dot{R})^{-\f{1}{2}} \ee
is a well-defined function on $H$ admitting a regular non vanishing
limit to $S_o$. We can thus conclude that $b^2$ vanishes at the same
rate as $\dot{R}$:\, $ b^2 \sim \dot{R} \, b_o^2$ as $v$ tends to
$v_o$. As an example, in the  Vaidya collapse, if one uses for $v$
the standard ingoing  Eddington-Finkelstein coordinate, then
$b_o=1/\sqrt{2}$. 

Let us return to the expression (\ref{q}) of the metric $q_{ab}$ on
$M$. Since $b$ vanishes as $v$ tends to $v_o$, and $V^a$ joins on
smoothly with $\bar\ell^a$ on $\Delta$ at $v = v_o$, it follows from
Eq. (\ref{V}) that $\rh^{\,a}$ diverges on $S_o$. On the other hand,
since
\be \rh_{\,a} = 2 b\, \partial_a v  \ee
on $H$, we conclude that $\rh_{\,a}$ vanishes at $S_o$. Finally, the
2 metric $\t{q}_{ab}$ smoothly approaches the intrinsic metric at
$S_o$.

Next, let us consider the expression (\ref{K}) of the extrinsic
curvature $K_{ab}$ of $H$. Since $V^a = 2b\, \rh^{\,a} = b (\ell^a -
n^a)$ on $H$, and $V^a$ joins on smoothly with $\bar\ell^a$ on
$\Delta$, it follows that we can smoothly extend $\bar\ell^a$ and
$\bar{n}^a$ from $\Delta$ to $H$ via
\be \lb^a:= b\, \ell^a , \quad\quad {\rm and}\quad\quad  \nb^a :=
b^{-1}n^a \, , \label{bar} \ee
(where we have used the fact that these null vector fields are
normalized via $\ell\cdot n = -2 = \bar\ell\cdot \bar{n}$.) Now the
part $S_{ab}$ of $K_{ab}$ in Eq. (\ref{K}) is related to the shear
tensors of these null vector fields:
\be S_{ab} = \frac{1}{2}(\sl_{ab} + \sn_{ab})\,
\equiv  \, \frac{1}{2}(\sl_{ab} + b
\snb)\, . \ee
Since $\nb^a$ is smooth on all of $M$, on $S_o$ we have $S_{ab} =
(1/2)\, \sl_{ab}$.  Now, on the DHs, we have the
following identity that arises directly from the constraint
equations on $H$ \cite{ak}:
\be \label{intHplus} \frac{1}{2G}= \oint_S\left[\frac{1}{16\pi
G}\left( |\sl|^2+2|\zeta|^2 \right)\,+\, T_{ab}\th^{\,
a} \l^b \right] d^2 V \ee
on any MTS $S$, where $\zeta^a$ is a vector field tangential to $S$,
given by Eq. (\ref{zeta}):
\be
 \zeta^a := \t{q}^{\,ab}\, \rh^{\,c}\,\grad_c\l_b= \t{\w}^a + \t{D}^a\ln |d R|\,
. \ee
Since each term in the integrand of (\ref{intHplus}) is positive
definite, by Taylor expanding the fields in $v$ we conclude that
$S_{ab}$ admits a regular limit to $S_o$.

Next, consider the term $\t\omega_a$ in the expansion (\ref{K}) of
$K_{ab}$. It is easy to check that $\t\omega_a =
-(1/2)\,\t{q}_a{}^b\, n^c\, \nabla_b \ell_c$. On the other hand we
also have the corresponding 1-form $\omega_a$ associated with the
barred null vectors $\lb^a, \nb^a$, namely $\omega_a =
-(1/2)\,\t{q}_a{}^b\, \nb^c\, \nabla_b \lb_c$, which is well-defined
on all of $M$. On $H$, the two are related by:
\be \label{relation} \omega_a = \t\omega_a + \t{D}_b \ln b\, . \ee
Recall, further, that $b = b_o\, \sqrt{\dot{R}}$ where $b_o$ has a
well-defined limit to $S_o$ which is nowhere zero. Since $\dot{R}$
is constant on any MTS $S$, on $H$ we can rewrite (\ref{relation})
as:
\be \omega_a = \t\omega_a + \t{D}_b \ln b_o \ee
which shows that $\t\omega_a$ admits a well-defined limit to $S_o$.

Finally, let us examine the coefficients $A$ and $B$ in the
expression (\ref{K}) of the extrinsic curvature. We have:
\be A = \f{1}{2}\, \t{q}^{ab}\nabla_a {\th_{\,b}} \quad {\rm and}
\quad B=\rh^{\,a} \rh^{\,b} \nabla_a \th_{\,b}. \ee
Writing $\th^{\,a}$ in terms of the null normals and using the fact that
each $S$ is a MTS, we find  $A\,=\,\tn/4 \,= \, b \tnb/4$, and so
$A \to 0$ as $v \to v_o$. To explore the limiting behavior of $B$,
let us rewrite it as
\be B=\frac{1}{2 b}\left( \kappa_V - V^a \partial_a\, \ln b \right),
\quad {\rm where} \quad \kappa_V := - \frac{1}{2}\,\nb_b V^a
\nabla_a V^b\, . \ee
Note that $\kappa_V$  is the surface gravity on DHs \cite{bf,ak}
which, at $S_o$, becomes the surface gravity $\kappa_{\lb}$ of the
IH which is positive because of our assumption that $\Delta$ is a
non-extremal IH. Thus, $\kappa_V$ has a well-defined limit to $S_o$.
However, because of the overall $1/b$ factor, for $B$ to have a
well-defined limit, $V^a \partial_a\, \ln b = \dot{b}/b$ must
approach $\kappa_{\lb} >0$ at a suitable rate. But this would imply
$b\, \sim \, \exp \kappa_{\lb}\, (v-v_o)$ as $v$ approaches $v_o$
which is impossible since $b =0$ on $S_o$. Thus, $B$ diverges in the
limit as the DH approaches equilibrium.

This concludes the discussion of the limiting behavior of $q_{ab}$
and $K_{ab}$ as $v \to v_o$ from below. In Eq. (\ref{q}),
$\rh_{\,a}$ tends to zero and $\t{q}_{ab}$ has a well-defined limit
which equals the intrinsic metric on $S_o$ induced by the IH
structure. In Eq. (\ref{K}), $A$ tends to zero, and $S_{ab}$ 
and $\t\omega_a$ have well-defined limits. However, $B$ diverges in
the limit. This implies in particular that the trace $K =
q^{ab}K_{ab}$ of the DH also diverges as we approach the isolated
horizon.

The divergence of $K$ has the following important consequence. Since
the dynamical horizons are space-like, one can use them as partial
Cauchy surfaces for the initial value problem of Einstein's
equations. If one could find a general solution to the constraint
equations for $(H, q_{ab}, K_{ab})$, one would have a
\emph{complete} description of all DHs that could ever arise in the
formation of a black hole. In the spherically symmetric case, thanks
to the systematic analysis of \cite{bi}, this problem has been
solved and the initial value equations have been reduced to a
single, second order linear `master equation'. As a result, one can
locally construct \emph{general} spherically symmetric space-times
admitting a DH and also locate the spherical DH in any given
spherically symmetric space-time \cite{bi}. It is tempting to try to
extend this analysis to general dynamical horizons. But because the
diffeomorphism and the Hamiltonian constraints are coupled in a
complicated fashion in the general setting, the standard strategy to
solve initial value constraints is to first decouple them by
assuming constancy of the trace $K$ of the extrinsic curvature
$K_{ab}$. However, because $K$ in fact diverges as one approaches
$S_o$, unfortunately this strategy cannot be used to solve the
initial value problem for general DHs that approach equilibrium. It
would be very interesting to devise another strategy by exploiting
the fact that the initial data we seek are very special, in that the
3-manifold $H$ admits a foliation by MTSs.

\subsection{Constraint equations}
\label{a1.2}

We will conclude our discussion of the behavior of fields on $H$ as
$H$ approaches equilibrium by listing a few consequences of the
field equations.

On the DH, by projecting the constraint equations along and
orthogonal to the MTSs and using $2\rh^{\,a} = \ell^a - n^a$ the
initial value equations can be written as:
\ba
2G_{ab}\th^{\,a}\l^b & = &16 \pi G T_{ab}\th^{\,a} \l^b \label{Gtl} ,\\
2G_{ab}\th^{\,a} n^b & = &16 \pi G T_{ab} n^a \l^b  \label{Gtn}, \\
G_{bc}\th^{\, b} \t{q}^c{}_a & = & 8 \pi G T_{bc}\th^{\,b}
\t{q}^c{}_a \label{Gtq}. \ea
Eq. (\ref{Gtl}) implies \cite{ak}
\be \R-|\sl|^2-2|\zeta|^2+2 \t{D}_a\zeta^a = 16 \pi G
T_{ab}\th^{\,a} \l^b \label{Hl}, \ee
and by integrating this equation on any MTS $S$ we obtain the Eq.
(\ref{intHplus}) which, as we have already noted, implies that
$\sl_{ab}$, $\zeta^a$ and $T_{ab}\th^{\,a} \l^b$ have well-defined
limits as we approach $v=v_o$ from below. Therefore $\R$ also has a
well-defined limit and, as one would expect, the limit is just the
scalar curvature of the 2-metric $\t{q}_{ab}$ on $S_o$. Furthermore,
(\ref{intHplus}) implies that the limiting values of $\sl_{ab}$,
$\zeta^a$ and $T_{ab}\th^{\,a} \l^b$ cannot all vanish. In fact, if
the IH $\Delta$ that $H$ approaches is generic in the precise sense
spelled out in \cite{abl1}  ---and this is in particular the case if
it is the Kerr IH--- then one can prove a stronger result:
$\sl_{ab}$ and $T_{ab}\,\ell^a \tau^b$ cannot both vanish
\cite{thesis}. On the other hand, the energy flux across any MTS is
dictated by these fields and that across any 2-sphere cross section
of $\Delta$ is zero. But there is no conflict because, even if
$\sl_{ab}$ and $T_{ab}\,\ell^a \tau^b$ cannot both vanish on $S_o$,
the energy flux across $S_o$ \emph{does vanish} because it is given
by \cite{ak}
\be \label{flux} E_{\rm flux}[S] = \oint_S |dR| \left[\frac{1}{16\pi
G}\left( |\sl|^2+2|\zeta|^2 \right)\,+\, T_{ab}\th^{\, a} \l^b
\right] d^2 V \ee
for any MTS $S$ and $|dR|$ vanishes in the limit.

Let us now turn to Eq. (\ref{Gtn}). By expressing the fields in
terms of those which have manifestly well-defined limits as $v\to
v_o$, we obtain
\be - \f{b^2}{2} \tnb^2 +V^a \partial_a\, \tnb + \kappa_V \tnb -
b^2|\snb|^2 -2 \t{D}_a \w^a-2|\w|^2+\R\, =\, 16 \pi G\,
T_{ab}\th^{\,a} n^b.\label{Hn} \ee
which reduces to
\be \mathcal{L}_{\lb}\,(\tnb)+ \kappa_{\lb} \tnb -2 \t{D}_a
\w^a-2|\w|^2+\R\, = \,8 \pi G T_{ab} \lb^a \nb^a, \ee
at $S_o$. This is precisely one of the field equations on the IH
side. Thus, the field equation under consideration is
\emph{automatically} continuous across the transition surface. It
does not further constrain the limiting behavior of geometrical
fields as the horizon attains equilibrium.

Finally let us examine the projection (\ref{Gtq}) of the vector
constraint into the MTSs. Again, we can express all fields in terms
of those which are manifestly smooth at $S_o$ to obtain
\be - \f{1}{4b}\, \t{D}_a(b^2\, \tnb) - \f{1}{2} b\tnb\, \zeta_a+
\f{1}{b}\, \t{D}^c(b S_{ac}) + \f{1}{2b}\, ({\cal L}_{V}\omega_a-\t{D}_a
\kappa_V)\, =\, 8 \pi G T_{bc}\th^{\,b} \t{q}^c{}_a . \ee
The limit of this equation is somewhat subtle since it contains
quotients of vanishing quantities. Moving these terms to the right
side and taking the limit, we obtain
\be \label{limmom} \f{1}{b_o}\, \t{D}^c(b_o\, \sl_{ac})|_{v=0}\,=\,
\lim_{v \to 0} \left\{ \f{1}{b}\, \left[ 8 \pi G T_{bc}\,\lb^b\,
\t{q}^c{}_a\, -\, ({\cal L}_{V}\w_a-\t{D}_a \kappa_V) \right] \right\}.
\ee
Since the left side is well-defined on $S_o$, we conclude the
numerator on the tight side must vanish in the limit. This is in
complete agreement with the equations on the IH side, which tell us
that each term in the numerator vanishes identically on the entire
IH. What we learn from Eq. (\ref{limmom}) is that the numerator on
the right side must vanish at a rate equal or faster than $b$.

We will conclude with an observation pertaining to the physically
most interesting case in which \emph{vacuum} equations hold on the
IH $\Delta$. Then, if the IH horizon is generic, our discussion of
Eq. (\ref{Gtl}) implies that $\sl_{ab}$ must be non-zero on $S_o$.
This in turn implies that the left side of (\ref{limmom}) is
necessarily nonzero.%
\footnote{For symmetric trace-free tensors $A_{ab}$ on $S$,
$\t{D}^bA_{ab}=0 \Rightarrow A_{ab}=0$. This follows from the fact that
there are no harmonic one-forms on 2-spheres. Since $A_{ab}$ can be
expanded as  $A_{ab}=\t{D}_a X_b+\t{D}_b X_a - (\t{D}_c X^c)\, q_{ab}$ for some
$X_a$, if $\t{D}^a A_{ab} =0$, we have $\t{D}^b A_{ab}= (d+ d^\dagger)^2
X_a=0$. This implies $X_a$ --and therefore $A_{ab}$-- must vanish.}
Therefore we conclude that $({\cal L}_{V}\w_a-\t{D}_a \kappa_V) $ goes
to zero as $\sqrt{\dot{R}}$. Consider now the case of a  $C^k$
transition, that is when $M$ is $C^{k+1}$ and the spacetime metric
is $C^{k}$. The vector field $\lb^a$ on $M$ is then $C^k$, which
implies that $({\cal L}_{V} \omega_a - \t{D}_a \kappa_V) $ is $C^{k-1}$.
Therefore, in local coordinates, it vanishes as $\sim v^k$ or
faster. Similarly, $b^2$ is $C^k$ and so $\dot{R} \sim v^n$ with $n
\geq k-1$. But the condition that the ratio in (\ref{limmom}) is
finite implies that actually $n \geq 2 k$. Thus $b$ and $R$ are
smoother than what one might initially expect.

\end{appendix}

\end{document}